\documentclass[aps,prd,twocolumn,groupedaddress,showpacs]{revtex4}
\usepackage{graphics}

\begin{document}


\boldmath
\title{Einstein's  equations 
from Einstein's inertial motion
\protect\\
and Newton's law for relative acceleration}
\unboldmath

\author{Christoph Schmid}     
\email{chschmid@itp.phys.ethz.ch}

\affiliation{ETH Zurich, Institute for Theoretical Physics, 
8093 Zurich, Switzerland}

\date{\today}

\begin{abstract}
We show that Einstein's equation
for  $R^{\hat{0} \hat{0}}(P)$
for nonrelativistic matter-sources in $P$
and for arbitrarily strong gravitational fields,
is identical with
Newton's equation
for the relative radial acceleration
of neighbouring freefalling test-particles, spherically averaged.--- 
With Einstein's concept of inertial motion 
($\equiv$ freefalling-nonrotating),
inertial worldlines ($\equiv$~geodesics) in Newtonian experiments
can intersect repeatedly. 
This is evidence for the space-time curvature encoded in $R^{\hat{0} \hat{0}}$.---
These two laws of Newton and Einstein 
are explicitely identical,
if one uses
(1)~our adapted space-time slicing 
     (generated by the radial 4-geodesics of the primary observer with
     worldline through $P$ and
     $\bar{u}_{\rm obs}  (P)= \bar{e}_{\hat{0}}  (P)$, 
(2)~our adapted Local Ortho-Normal Bases, LONBs 
(radially parallel with the primary observer's LONB),
and 
(3)~Riemann normal 3-coordinates  (centered at the primary observer). 
Hats on indices denote LONB components.---
Our result:
Full general relativity follows 
from Newton's law of relative acceleration 
by using Lorentz covariance and energy-momentum conservation
combined with the Bianchi identity.---
The gravitational field equation of 
Newton-Gauss  
and Einstein's field equation for
$R^{\hat{0} \hat{0}} (P)$ are both 
      linear 
in gravitational fields,
if the primary observer ($\equiv$ worldline through $P$) is 
      inertial.

      Einstein's principle of equivalence between fictitious forces 
and gravitational forces is formulated as a precise equivalence theorem 
with explicit equations of motion from  general relativity.
With this equivalence theorem,
the gravitational field equation of 19th-century
Newton-Gauss physics and Einstein's field equation for
$R^{\hat{0} \hat{0}} (P)$ are both  
       bilinear in the gravitational forces
for 
       non-inertial
 primary observers.

       $R^{\hat{0} \hat{0}} = - \mbox{div} \vec{E}_{\rm g} \, $ and
$\, R^{\hat{i} \hat{0}} = - (\mbox{curl} \vec{B}_{\rm g}/2)^{\hat{i}} \,$
hold exactly in general relativity for inertial primary observers,
if one uses our space-time slicing and our LONB's.
The gravitoelectric $\vec{E}_{\rm g}$ and the gravitomagnetic
$\vec{B}_{\rm g}$ fields are defined and measured exactly with
nonrelativistic test-particles via $(d/dt) (p_{\hat{i}})$ and
$(d/dt) (S_{\hat{i}})$ in direct correspondence with the
electromagnetic $(\vec{E},  \vec{B})$ fields.
The $(\vec{E}_{\rm g},  \vec{B}_{\rm g})$ fields are identical with
the Ricci connection for any observer's LONBs and for displacements along any 
observer's worldline, $(\omega_{\hat{a} \hat{b}})_{\hat{0}}$.

In the explicit equations of particle-motion of general relativity
(using our adapted space-time slicing and our adapted LONBs),
there are precisely two gravitational forces 
equivalent to the two fictitious forces  on the worldline of the observer:
the force from $ \vec{E}_{\rm g},  $ 
equivalent to the fictitious force measured by an accelerated observer, 
and the force  from $ \vec{B}_{\rm g},  $ 
equivalent to the ficitious Coriolis force 
for a rotating observer.

      The exact Ricci curvature component
$R^{\hat{i} \hat{0}}$ 
can be measured with non-relativistic test particles. 
$(R^{\hat{i} \hat{0}},  \, R^{\hat{0} \hat{0}})_P$ 
are linear in the gravitational fields $(\vec{E}_{\rm g}, \, \vec{B}_{\rm g})$
for inertial primary
observers with worldlines through $P$.
For non-inertial primary observers,
$(R^{\hat{i} \hat{0}}, \, R^{\hat{0} \hat{0}})$ 
are bilinear in
the gravitational fields $(\vec{E}_{\rm g},  \, \vec{B}_{\rm g})$, 
which are the only gravitational fields 
in $(R^{\hat{i} \hat{0}}, \, R^{\hat{0} \hat{0}}).$

      Einstein's $R^{\hat{0} \hat{0}}$ equation for non-relativistic
matter and for inertial primary observers gives the Gauss law,
$\mbox{div} \vec{E}_{\rm g} = - 4 \pi G_{\rm N} \rho_{\rm mass}$,
and Einstein's $R^{\hat{0} \hat{i}}$ equation gives the
gravito-magnetic Amp\`ere law, $\mbox{curl} \vec{B}_{\rm g} =
- 16 \pi G_{\rm N} \vec{J}_{\rm mass}$.---
For relativistic matter and inertial primary observers, Einstein's
$R^{\hat{0} \hat{0}}$ equation gives $\mbox{div} \, \vec{E}_{\rm g} =
- 4\pi G_{\rm N} (\tilde{\rho}_{\rm \varepsilon} + 3 \tilde{p})$,
where
$\tilde{\rho}_{\varepsilon}$ is the energy density, and
$(3 \tilde{p})$ is the trace of the momentum-flow 3-tensor,
both in the frame with $\bar{u}_{\rm obs} = \bar{e}_{\hat{0}}$.
Einstein's $R^{\hat{0} \hat{i}}$ equation for an inertial observer
gives the relativistic
gravito-magnetic Amp\`ere law, $\mbox{curl} \vec{B}_{\rm g} =
-16 \pi G_{\rm N} \vec{J}_{\varepsilon}$.

      The remaining six Ricci components, $R^{\hat{i} \hat{j}}$,
involve the curvature of space-space plaquettes, which are unmeasurable
with non-relativistic particles in quasi-local experiments.

With our primary-observer-adapted spacetime splitting and LONBs,
the equations of motion of general relativity
for particles without nongravitational forces 
and for a noninertial primary observer are 
(1)  
form-identical with the 19th-century equations of Newtonian mechanics
for nonrelativistic particles, 
(2) 
form-identical with the equations of motion for special relativity with the 
obvious replacements 
$(\vec{E}, \, \vec{B}) \Rightarrow  
(\vec{E}_{\rm g}, \, \vec{B}_{\rm g}) $ and 
$q \Rightarrow \varepsilon \equiv$ total energy of particle.---
We formulate the precise theorem of equivalence 
of fictitious and gravitational
forces in the equations of motion.

\end{abstract}
%


\maketitle



\boldmath
\section{Gravito-electric and gravito-magnetic fields
$\vec{E}_{\rm g}, \, \vec{B}_{\rm g}$}
\unboldmath
\label{grav.el.magn.fields}

Our {\it exact} and {\it general} operational definition 
in  arbitray spacetimes
of the gravitoelectric field  $\vec{E}_{\rm g}$ 
       measured by any observer
is probably new.

In  contrast to the literature, 
we use
no perturbation theory on a background geometry, 
no ``weak gravitational fields'',
no ``Newtonian limit''. 

For an observer with his 
Local Ortho-Normal Bases (LONBs) on his worldline
our exact operational definition
of the gravitoelectric field $\vec{E}_{\rm g}$
is by measuring the acceleration of 
{\it quasistatic} freefalling test-particles 
in analogy to the operational definition of the ordinary electric field.
We must replace 
the particle's charge $q$ 
by its rest mass $m$.
This gives the operational definition of the gravito-electric field, 
\begin{eqnarray}
&& \quad \quad m^{-1} \, \frac{d}{dt} \, \, p_{\hat{i}} \, \, \,
\equiv \, \, E_{\hat{i}}^{\, (\rm g)} 
\nonumber
\\
&& \quad \quad \Leftrightarrow  \quad \quad \vec{a}_{\rm ff} 
\, \,  \,  =   \,  \, \vec{E}^{\, (\rm g)}  \, \, =  \, \, \vec{g}, 
\label{def.E}
\\
&&\mbox{for freefalling, quasistatic test particles.}
\nonumber 
\end{eqnarray}
%
The local time-interval  $dt$ is measured on the observer's wristwatch.
The measured 3-momentum is $p_{\hat{i}}$,
and
the measured acceleration of the quasistatic particle
relative to the observer 
is $\vec{a}_{\rm ff} = \vec{g} \equiv \, $ 
gravitational acceleration.

We use use the  method of  \'Elie Cartan,  
who gives vectors $\bar{V}$ (and tensors) at any spacetime point~$P$
by their components  $V^{\hat{a}}$
in the chosen Local Ortho-Normal Basis.~---
The LONB-components  $\, V^{\hat{a}} \, $  are {\it directly measurable}.
This is in stark contrast to coordinate-basis components $V^{\alpha},$
which are not measurable before one has obtained $g_{\alpha \beta}$
by solving Einstein's equations for the specific problem at hand.~---
Cartan's method uses coordinates only  
in the mapping 
from an event $P$ to the event-coordinates,
$P \Rightarrow x^{\mu}_P.$ 

It is  important to distinguish LONB-components,
denoted by hats,
from coordinate-basis components,  
denoted without hats
in the notation of  
Misner, Thorne, and Wheeler \cite{MTW}.
We denote spacetime LONB-indices by $(\hat{a}, \hat{b}, ...)$, 
3-space LONB-indices by $(\hat{i}, \hat{j}, ...)$, and 
coordinate-indices by Greek letters.

LONBs {\it off} the observer's worldline are not needed in
   Eq.~(\ref{def.E}),
because a particle 
released from rest 
will still be on the observer's
worldline after an infinitesimal time $\delta t$,
since $\delta s \propto (\delta t)^2 \Rightarrow 0, $
while  $\delta v \propto \delta t \neq 0$.

{\it Arbitrarily strong} gravito-electric fields $\vec{E}_{\rm g}$ 
of general relativity can be
measured {\it exactly} with freefalling test-particles
which are {\it quasistatic} relative to the observer
in Galilei-type experiments,
      Eq.~(\ref{def.E}).~---
But this same  measured $\vec{E}_{\rm g}$ 
is {\it exactly} valid for {\it relativistic} test-particles
in the equations of motion of general relativity,
   Eqs.~(\ref{eqs.motion.p.noninertial.2}), and in the field equations.

For the gravitomagnetic field  $\vec{B}_{\rm g}$, 
we give the
 {\it exact operational definition} 
in  arbitray spacetimes,
which is 
analogous to the modern definition of the ordinary magnetic field:
$\, \vec{B}_{\rm g} $ is defined via the gravitational torque
causing the spin-precession of 
quasistatic freefalling test particles
with spin $\vec{S}$ and free of nongravitational torques 
(or by quasistatic gyroscopes).
In the gyro-magnetic ratio $q/(2m)$ of electromagnetism, 
the particle charge $q$ must be replaced by the rest-mass $m$, 
therefore the gyro-gravitomagnetic ratio is (1/2),
%
\begin{eqnarray}
\frac{d}{dt} \, S_{\, \hat{i}} 
\, \, \, &\equiv&   \, \,
 [\, \vec{S} \, \times  \, (\vec{B}_{\rm g}/2) \,]_{\, \hat{i}}
\label{def.B}
\\ 
\Leftrightarrow \, \,  \vec{\Omega}_{\rm gyro} 
\, \,  \,  &=&   \,   - \, (\vec{B}_{\rm g}/2), 
\quad \mbox{quasistatic gyroscope},
\label{Omega.vs.B}
\\ 
1/2 \, \, \, \, \, &=& \, \, \mbox{gyro-gravitomagnetic ratio}.
\nonumber
\end{eqnarray}
%

The gravitomagnetic field    $\vec{B}_{\rm g}$
can also be measured 
by the deflection of  freefalling test particles,  
nonrelativistic relative to the observer,
\begin{eqnarray}
&& \frac{d}{dt} \, \vec{p} 
\, = \,  m \, [ \, \vec{E}_{\rm g} \, 
+ \, (\vec{v} \times \vec{B}_{\rm g}) \, ].
\label{Coriolis.force}
\end{eqnarray}
%
The second term, 
$ \vec{F}_{\rm g} = m   (\vec{v} \times \vec{B}_{\rm g})$
is the Coriolis force of Newtonian physics, 
which arises in rotating reference frames.
It has the same form as the Lorentz force 
of electromagnetism with the
charge $q$ replaced by the rest mass $m$ for a nonrelativistic particle.~---
Our definition of $\vec{B}_{\rm g}$ agrees with Thorne et al \cite{Thorne}.

The gravitomagnetic field $\vec{B}_{\rm g}$ 
of general relativity
has been measured 
by Foucault
with 
gyroscopes precessing 
relative to his LONBs, 
$\bar{e}_{\hat{0}} = \bar{u}_{\rm obs} \,$ and  $\, \bar{e}_{\hat{i}} =$ 
(East, North, vertical).
Recently, $ \vec{B}_{\rm g}$ has been measured 
on {\it Gravity Probe B} 
by gyroscope precession
relative to  LONBs 
given by the line of sight to quasars
resp. 
distant stars without measurable proper motion.

In 1893, Oliver Heaviside, 
in his paper {\it A Gravitational and Electromagnetic Analogy}
    \cite{Heaviside}, 
gave the same operational definition 
of the gravito-electric field  $\vec{E}_{\rm g},$
and he postulated the gravito-magnetic field $\vec{B}_{\rm g}$
in analogy to Amp\`ere-Maxwell.   

For an inertial observer (freefalling-nonrotating)
with worldline through $P$, 
$(\vec{E}_{\rm g}, \, \vec{B}_{\rm g})_P$ are zero.


{\it Arbitrarily strong} gravito-magnetic field $\vec{B}_{\rm g} \, $   
of general relativity 
can be measured  {\it exactly} 
with the precession of {\it quasistatic} 
gyroscopes 
   Eq.~(\ref{def.B}),
or with the Coriolis-deflection of {\it nonrelativistic} test particles,
   Eq.~(\ref{Coriolis.force}).~---
But this same  measured $\vec{B}_{\rm g}$ 
is {\it exactly} valid 
for {\it relativistic} test-particles
in the equations of motion,
   Eqs.~(\ref{eqs.motion.p.noninertial.2}, 
         \ref{Fermi.transport.eqs.motion.p.noninertial.2}),
and in the field equations.

\newpage

\subsection{Ricci's LONB-connection}
\label{def.Ricci.coeff}

Relative to an 
airplane on the shortest path (geodesic)
from Zurich to Chicago, 
the Local Ortho-Normal Bases (LONBs), chosen to be in the directions 
``East'' and ``North'',
otate relative to the geodesic (relative to parallel transport) 
with a rotation angle $\delta \alpha$ per
path length $\delta s,$ 
i.e. with the rotation rate $ \, \omega = (d\alpha/ds).$

For an infinitesimal displacement $\, \delta \vec{D}  \,$
along a geodesic in any direction, 
the infinitesimal rotation angle $\, \delta \alpha  \, $ of LONBs
is given by a linear map encoded by the Ricci rotation coefficients
$\, \omega_{\hat{c}}  \,$
%
\begin{eqnarray}
\delta \alpha \, &=& \, \omega_{\hat{c}} \, \, \delta D_{\hat{c}}.
\nonumber
\end{eqnarray}
The Ricci rotation coefficients are also called connection coefficients,
because they connect the LONBs at infinitesimally neighboring points by a 
rotation relative to
the infinitesimal geodesic between these points 
(i.e. relative to parallel transport).

For non-geodesic displacement curves,   
the tangents for infinitesimal displacements
are infinitesimal geodesics.
For the connection coefficients, 
it is irrelevant, 
whether the displacement curve is geodesic or non-geodesic,
only the tangent vectors matter, 
either the coordinate basisvector, 
$\partial_{\gamma} = \bar{e}_{\gamma}, \, $ 
or   the LONB-vector, $ \, \partial_{\hat{c}} = \bar{e}_{\hat{c}}.$

For  Ricci connections (LONB rotations), it is irrelevant, whether we are in
curved space or e.g. in the Euclidean plane with polar coordinates 
$(r, \, \phi)$ and LONBs $(\vec{e}_{\hat{r}}, \, \vec{e}_{\hat{\phi}})$,
where for a displacement vector $\vec{e}_{\hat{\phi}}$ 
the LONB-rotation angle is
$(\omega_{\hat{r} \hat{\phi}})_{\hat{\phi}} = 1/r$.

The  Ricci rotation coefficients 
$\omega_{\hat{c}}$
are directly {\it measurable} because of their 
{\it LONB}-displacement-index.~---
But for computations of curvature
   in Sect.~\ref{sect.Cartan.LONB.rot.closed.curve}, 
a line-integral along a displacement curve
$ {{\cal C}} \, $
calls for infinitesimal displacement vectors
with contravariant components  $\, \delta x^{\gamma}_{PP'},  \,$
the infinitesimal difference of coordinates $x^{\gamma}_P$.
This calls for a 
linear map from infinitesimal coordinate-displacement 
vectors  $\, \delta x^{\gamma}_{PP'}  \,$ (input)
to the corresponding measured 
LONB-rotation angles $\, \delta \alpha \,$ (output),
%
\begin{eqnarray}
\delta \alpha \, &=& \, \omega_{\gamma} \, \, \delta x^{\gamma}.
\nonumber
\end{eqnarray}
%
This linear map is the LONB-connection 
1~-~form $\tilde{\omega}, \, $
given explicitely by covariant components  $\, \omega_{\gamma} \, $
($\equiv$ 1-form components). 
In this paper, Greek indices always refer to a coordinate basis.
Using 1-form components
makes the line-integral free of metric factors
$\, g_{\mu \nu},$
%
\begin{eqnarray}
\alpha ({\cal C})   
&=&  \int_{{\cal C}}  \omega_{\gamma} \, \, dx^{\gamma}.
\label{line.integral.omega}
\end{eqnarray}
%
A line-integral with $dx^{\gamma}$ cries out to have some 1-form 
$\, \sigma_{\gamma} \, $ as an integrand.
Vice versa,
the connection 1-form $\, \omega_{\gamma} \, $ 
cries out to be contracted with a 
contravariant displacement vector to give a rotation angle,
$\,\delta  \alpha = \omega_{\gamma} \, \delta x^{\gamma}. \, $
%
%
This is index-matching.

The rotation of the chosen LONBs 
$\, (\vec{e}_{\hat{x}}, \, \vec{e}_{\hat{y}})   \,$ 
relative to the geodesic from $P$ to $Q$ 
(i.e. relative to parallel transport)
is given by 
the rotation matrix,
\[
\left(
\begin{array}
 { c}
\vec{e}_{\hat{x}}
 \\
\vec{e}_{\hat{y}}
\end{array}
\right)_Q 
\, \, = \, \, 
\left(
\begin{array}
 {*{1}{c@{\: \, \,  \:}} c}
\, \cos  \alpha & \sin  \alpha  \\
-  \sin  \alpha & \cos  \alpha
\end{array}
\right) \,
\left(
\begin{array}
 { c}
\vec{e}_{\hat{x}}\\
\vec{e}_{\hat{y}}
\label{rotation}
\end{array}
\right)_P. 
\]
%
For infinitesimal displacements, the {\it infinitesimal rotation matrix} is,
%
\[
\left(
\begin{array}
 { c}
\vec{e}_{\hat{x}}
 \\
\vec{e}_{\hat{y}}
\end{array}
\right)_Q  \, \, = \, \, 
\left[ \, 1  \, + \, \alpha \, \left(
\begin{array}
 {*{1}{c@{\: \, \,  \:}} c}
\, 0 & 1  \\
-  1 & 0
\end{array}
\right)
\right] \, 
\left(
\begin{array}
 { c}
\vec{e}_{\hat{x}}\\
\vec{e}_{\hat{y}}
\label{Lorentz.trsf}
\end{array}
\right)_P. 
\]
The infinitesimal LONB-rotation matrix  
$\, \delta R_{\hat{i} \hat{j}} \, $ is given by  the linear map
from the infinitesimal coordinate-displacement vector $D^{\gamma}$,
\begin{eqnarray}
\delta R_{\hat{i} \hat{j}} 
&=&    (\omega_{\hat{i} \hat{j}})_{\gamma} \, \, \, 
\delta D^{\gamma},
\label{inf.rotation}
\\
  \omega_{\hat{1} \hat{2}} 
\, = \, - \, \omega_{\hat{2}  \hat{1}}
&=& \alpha_{\hat{1} \hat{2}} \, =  \, 
\mbox{rotation angle in} \, [\, \hat{1}, \, \hat{2} \, ] \, 
\mbox{plane}.  
\nonumber
\end{eqnarray}
The coefficients    $\, (\omega_{\hat{i} \hat{j}})_{\gamma} \, $
are the
{\it connection 1-form components.}

The rotational change  of a LONB vector 
relative to parallel transport 
under a displacement in the coordinate $x^{\gamma}$ is
called the covariant derivative of  $\vec{e}_{\hat{i}}$ 
with respect to the coordinate $x^{\gamma}$
and denoted with the symbol $\nabla_{\gamma},$ 
\begin{eqnarray}
(\partial \vec{e}_{\hat{i}} / \partial x^{\gamma})_{\rm relat.to \, parall.trsp.}
\, &\equiv& \, 
\nabla_{\gamma} \, \vec{e}_{\hat{i}}.
\label{covariant.derivative}
\end{eqnarray}
Combining Eqs.~(\ref{inf.rotation})
and~(\ref{covariant.derivative})
gives,
\begin{eqnarray}
\nabla_{\gamma} \, \vec{e}_{\hat{j}} \, &=& \,
\vec{e}_{\hat{i}} \,
(\omega_{\hat{i} \hat{j}})_{\gamma}.
\label{covar.deriv.of.LONB}                    
\end{eqnarray}

In 3-space, for an infinitesimal displacement from $P$ to $Q$,
the LONBs $\vec{e}_{\hat{i}}$ rotate  relative to parallel transport,
which can be given 
by (1) the  tangent vector to the geodesic curve
and (2) the spin-axes of two transported gyroscopes.  
The infinitesimal rotations are given by 
the antisymmetric matrix with components
$(\omega_{\hat{1} \hat{2}})_{\gamma}, \, \,
(\omega_{\hat{2} \hat{3}})_{\gamma}, \, \,
(\omega_{\hat{3} \hat{1}})_{\gamma}.$

In curved (1+1)-spacetime, 
the Lorentz transformation 
of the chosen LONBs 
relative to a given displacement geodesic
is a Lorentz boost
$\, L^{\hat{a}}_{\, \, \, \hat{b}}, \, $ 
\[
\left(
\begin{array}
 { c}
\bar{e}_{\hat{t}}
 \\
\bar{e}_{\hat{x}}
\end{array}
\right)_Q 
\, \, = \, \, 
\left(
\begin{array}
 {*{1}{c@{\: \, \,  \:}} c}
  \cosh  \chi & \sinh  \chi  \\
  \sinh  \chi & \cosh  \chi
\end{array}
\right) \,
\left(
\begin{array}
 { c}
\bar{e}_{\hat{t}}\\
\bar{e}_{\hat{x}}
\label{L.boost}
\end{array}
\right)_P, 
\]
%
\begin{eqnarray}
\mbox{rapidity}  &\equiv&  \chi \, \, = \, \, \mbox{additive,}
\quad \quad \, \, \, \, 
\tanh \chi \, \equiv \, v/c. \, \, 
\nonumber
\end{eqnarray}
%
%
For infinitesimal displacements,
the infinitesimal Lorentz boost $\, L^{\hat{a}}_{\, \, \, \hat{b}}, \, $ is,
%
\[
\left(
\begin{array}
 { c}
\bar{e}_{\hat{t}}
 \\
\bar{e}_{\hat{x}}
\end{array}
\right)_Q  \, \, = \, \, 
\left[ \, 1 \, + \,  \chi \, \left(
\begin{array}
 {*{1}{c@{\: \, \,  \:}} c}
  0 & 1  \\
  1 & 0
\end{array}
\right)                                                             
\right] \, 
\left(
\begin{array}
 { c}
\bar{e}_{\hat{t}}\\
\bar{e}_{\hat{x}}
\nonumber
\end{array}
\right)_P. 
\]
%
In space-time, we denote vectors by a bar, $\bar{V}$,
in 3-space, we denote vectors by an arrow, $\vec{V}.$

In curved (3+1)-spacetime, and
with {\it two lower indices,}
   $\, \omega_{\hat{a} \hat{b}} \,$ 
is {\it antisymmetric} for Lorentz boosts and for rotations, 
 %
\begin{eqnarray}     
\delta L_{\hat{a}\hat{b}} \, 
= \,   (\omega_{\hat{a} \hat{b}})_{\gamma} \, \, \, \delta D^{\, \gamma},
\, \, \,  && \, \, \,  
\omega_{\hat{i} \hat{0}} \, = \, - \, \omega_{\hat{0} \hat{i}}
\,  = \,   \chi_{\hat{i} \hat{0}}.
\nonumber
\end{eqnarray}
%


For a displacement in observer-time,
the {\it exact} Ricci connection coefficients 
  $ (\omega_{\hat{a}   \hat{b}})_{\hat{0}}$
of general relativity 
can be {\it measured} 
in {\it quasistatic} experiments.
But these Ricci connection coefficients 
{\it predict} the motion of {\it relativistic} particles
with the equations of motion,   
   Eqs.~(\ref{eqs.motion.p.noninertial.2}, 
         \ref{Fermi.transport.eqs.motion.p.noninertial.2}).


\boldmath
\subsection{$(\vec{E}_{\rm g}, \, \vec{B}_{\rm g}) \, $
identical with 
\protect\\
Ricci connection in time 
$\, (\omega_{\hat{a} \hat{b}})_{\hat{0}}$}
\label{sect.E.B.Ricci.2}
\unboldmath

Our gravitoelectric field $\, \vec{E}_{\rm g}$
is identically equal to 
the negative of the Ricci Lorentz-boost coefficients 
for a displacement in time,
\begin{eqnarray}
E^{(\rm g)}_{\hat{i}} 
 \, &=& \,  - (\omega_{\hat{i}  \,   \hat{0}})_{\hat{0}}.
\label{E.Ricci}
\end{eqnarray}
The proof is immediate, see the next equation:
from the point of view of the observer with his LONBs along his worldline,
the gravitational acceleration~$g_{\hat{i}} = a_{\hat{i}}^{(\rm ff \, particle)}$
of freefalling {\it quasistatic} test-particles 
(starting on the observer's worldline)
is by definition identical to  the 
{\it exact} gravitoelectric field~$E_{\hat{i}}$ of general relativity
from 
  Eq.~(\ref{def.E}).~---
But from the point of view of freefalling test-particles,
the acceleration of the quasistatic observer with his LONBs
is by definition identical to 
the {\it exact} Ricci LONB-Lorentz-boost coefficients 
   $\, (\omega_{\hat{i} \hat{0}})_{\hat{0}}$,
\begin{eqnarray}
E^{(\rm g)}_{\hat{i}} 
 \, &\equiv& \, 
[(a_{\hat{i}})^{\, (\rm ff \, particle)}_{\, \rm relat.to \, obs.}]_{\rm quasistatic}
\nonumber
\\ 
&=&  \,  - \, [(a_{\hat{i}})^{\, (\rm obs.)}_{\, \rm relat.to \, ff}]_{\rm quasistatic}
\, \, \equiv \, \, - \, (\omega_{\hat{i} \hat{0}})_{\hat{0}}.
\label{E.2acc.Ricci}
\end{eqnarray}
%

The {\it exact} Ricci connection coefficients
  $ (\omega_{\hat{i}   \hat{0}})_{\hat{0}}$
of general relativity in arbitrarily {\it strong gravitational fields} 
can be directly {\it measured} 
in {\it quasistatic} experiments 
by the acceleration of freefalling test particles 
relative to the LONBs of the observer as Galilei did.

For the gravitomagnetic 3-vector field $\vec{B}^{(\rm g)}$, we introduce its
Hodge-dual antisymmetric 2-tensor $B^{(\rm g)}_{\hat{j}\hat{k}}$,
\begin{eqnarray}
B^{(\rm g)}_{\hat{i} \hat{j}} \, &\equiv& \, 
\varepsilon_{\hat{i} \hat{j} \hat{k}} \, B_{\hat{k}}^{(\rm g)},
\end{eqnarray}
where $\varepsilon_{\hat{i} \hat{j} \hat{k}}$
is the Levi-Civita tensor, totally antisymmetric,
whose primary definition is given in a LONB, 
%
\begin{eqnarray}
\varepsilon_{\hat{1} \hat{2} \hat{3}}   &\equiv&   + \, 1 \, \, \,
\mbox{for LONB with  positive orientation}, \, \, 
\label{Levi.Civita}
\\
B_{\hat{1} \hat{2}} &\equiv& B_{\hat{3}} \, \, \, 
\mbox{and cyclic permutations.}
\nonumber
\end{eqnarray}
The Levi-Civita tensor in a coordinate-basis in 3 dimensions, 
$\varepsilon_{\alpha \beta \gamma}$, cannot be known before Einstein's
equations have been solved for the specific problem at hand.

Recall from 
   Eq.~(\ref{Omega.vs.B})
that the gyro-precession relative to the observer
equals $(- \vec{B}_{\rm g}/2)$, hence
\begin{eqnarray}
 (- \, B^{(\rm g)}_{\hat{i} \, \hat{j}}/2) 
 \, &\equiv&  \, 
[(\Omega_{\hat{i} \hat{j}})^{(\rm gyro)}_{\rm relat. to \, obs.}]_{\rm quasistatic} 
\nonumber
\\
&=& \, - \, [(\Omega_{\hat{i} \hat{j}})^{(\rm obs)}_{\rm relat. to \,
    gyro}]_{\rm quasistatic} 
\, \, \equiv \, \,  
- \, (\omega_{\hat{i}  \,   \hat{j}})_{\hat{0}}.
\nonumber
\\
&&
\label{B.Ricci}
\end{eqnarray}
The exact Ricci connection coefficients
  $ (\omega_{\hat{i}   \hat{j}})_{\hat{0}}$
of general relativity 
can be {\it directly measured} 
in {\it quasistatic} experiments 
by the precession of gyroscopes
relative to the LONBs of the observer as  Foucault did.


All Ricci connection coefficients 
for a displacement in time,
$\, (\omega_{\hat{a} \hat{b}})_{\hat{0}}, \, $  
which have all indices in LONBs,
can be directly and 
{\it  exactly} measured 
in arbitrarily {\it strong gravitational fields} of general relativity
using freefalling test particles and gyroscopes
which are {\it quasistatic} relative to the observer 
with $\bar{u}_{\rm obs} = \bar{e}_{\hat{0}}$ 
in Galilei-type and Foucault-type experiments.
Therefore, it is superfluous to use relativistic test-particles to
measure 
the Ricci connection coefficients for a displacement in time.

Cartan's LONB-connection 1-form 
 $(\omega^{\hat{a}}_{\, \, \,  \hat{b}})_{0}$, 
with its displacement 
index in the
coordinate basis, 
are measurable as soon as a coordinatization
$\, P \Rightarrow x^{\mu}_P \,$ is chosen.
No metric coefficients $g_{\alpha\beta}$ are needed.

In striking contrast, the connection coefficients for  coordinate bases, 
the Christoffel symbols, 
$\, (\Gamma^{\alpha}_{\, \, \,  \beta})_{\gamma} 
\equiv \Gamma^{\alpha}_{\, \,  \, \beta\gamma}, \, $
have no direct physical-geometric meaning,
they are not measurable until the metric fields $\, g_{\mu \nu} (x) \,$
have  been obtained 
by  either {\it solving Einstein's equations}
or going out and 
measuring distances, time-intervals, angles, and Lorentz-boost-angles
in a fine-mesh coordinate grid.~---
We write Christoffel connection-1-form coefficients 
with a bracket:  
$\Gamma^{\alpha}_{\, \, \,  \beta \gamma} \, \equiv \, 
(\Gamma^{\alpha}_{\, \, \,   \beta})_{\gamma}$. 
Inside the bracket are the coordinate-basis transformation-indices 
   $\, (\alpha, \, \beta), \, $
outside the bracket is the coordinate-displacement index $\, \gamma$.

Conclusion:
\begin{itemize}
\item {\it Ricci} connection coefficients 
      $\, (\omega^{\hat{a}}_{\, \, \, \hat{b}})_{\hat{d}} \, $
      are {\it directly measurable} for given LONBs,
\item {\it Cartan} connection coefficients
        $\, (\omega^{\hat{a}}_{\, \, \, \hat{b}})_{\mu} \, $
      are {\it measurable} for given LONBs
      after a coordinatization-mapping $\, P \Rightarrow x^{\mu}$
      is chosen, but  no metric is needed,
\item {\it Christoffel} connection coefficients 
      $\,  (\Gamma^{\alpha}_{\, \, \, \beta})_{\delta} 
   \equiv  \Gamma^{\alpha}_{\, \, \, \beta \nu} \, $
      {\it cannot be known}
      until one has solved Einstein's equations for
      the problem at hand in order to obtain 
      the metric $\, g_{\alpha \beta} \,$ for the chosen coordinate system.
\end{itemize}

The connection-coefficients for local ortho-normal bases (Ricci and Cartan)
are  more efficient than Christoffel's connection-1-form coefficients: 
In two spatial dimensions,
Ricci and Cartan connections each need  only {\it one rotation angle} 
for a given  
displacement.
In contrast,  the Christoffel-connection 1-form 
      $\, (\Gamma^{\alpha}_{\, \, \,  \beta})_{\gamma} \, $
for a given displacement 
needs {\it four numbers} for the coordinate-basis transformation, 
two for stretching/compressing basis vectors, one for skewing them, 
and one for rotating
them. 

\newpage

\subsection{Equations of motion with
\protect\\
general LONBs and coordinates}

We  obtain 
the equations of motion in terms of Ricci coefficients
from (1) the equations of motion 
in terms of $\, (\vec{E}_{\rm g}, \, \vec{B}_{\rm g} )   \,$
for {\it quasistatic freefalling-nonrotating} test-particles, 
     Sect.~\ref{grav.el.magn.fields}, 
(2) the identity
of 
$(E^{(\rm g)}_{\hat{i}}, B^{(\rm g)}_{\hat{i} \hat{j}})$
with 
the Ricci connection for a displacement in time,
$(\omega_{\hat{a} \hat{b}})_{\hat{0}},$   
  Sect.~\ref{sect.E.B.Ricci.2},
%
\begin{eqnarray}
 m^{-1} \, \frac{d}{d\hat{t}} \, \, p_{\hat{i}}
\, + \, 
(\omega_{\hat{i} \hat{0}})_{\hat{0}} 
 \, &=& \, 0, \quad \quad    i = 1,2,3,  
\nonumber
\\
\frac{d}{d\hat{t}} \, \, S_{\hat{i}} 
 \, + \,  (\omega_{\hat{i}  \hat{j}})_{\hat{0}} \, \, S_{\hat{j}}
 \, &=&  \, 0.
\label{nonrelat.geodesic.Fermi.spin.transport} 
\end{eqnarray} 
%

At the level of first derivatives of vectors, 
there cannot be curvature effects, 
we are at the level of special relativity,
and the Lorentz-covariant extension of 
   Eqs.~(\ref{nonrelat.geodesic.Fermi.spin.transport}) is trivial:
we must include
spatial displacements in $(dx^{\mu}/dt) \, $
and spatial components of the vector $p^{\hat{b}}. \,$
This Lorentz-covariant extension of 
   Eq.~(\ref{nonrelat.geodesic.Fermi.spin.transport})
gives
the relativistic {\it geodesic equation} 
and  Fermi {\it spin transport equation},
\begin{eqnarray}
\frac{d}{dt} \, \, p^{\hat{a}}
\, \, + \, \, 
(\omega^{\hat{a}}_{\, \, \,  \hat{b}})_{\mu} \, \, \,  
p^{\hat{b}} \, \, \, \frac{dx^{\mu}}{dt} 
 \, \, &=& \, \, 0,
\nonumber
\\ 
\frac{d}{dt} \, \, S^{\hat{a}}
\, + \, 
(\omega^{\hat{a}}_{\, \, \,  \hat{b}})_{\mu} \, \, \, 
S^{\hat{b}} \, \, \, \frac{dx^{\mu}}{dt} 
 \, &=& \, 0.
\label{geodesic.Fermi.spin.transport.eq}
\end{eqnarray}
%

The covariant derivative $\nabla_{\mu}$ of vector- and tensor-fields
can be defined by 
\begin{eqnarray}
&&\quad \quad \, \, \, (\nabla_{\mu} \bar{V})_{\hat{a}} \, \, \equiv \, \,  
\partial_{\mu} (\bar{V}_{\hat{a}})
\nonumber
\\
&&\mbox{for LONBs 
parallel in direction} \, \partial_{\mu}.
\end{eqnarray}
%

The covariant derivative along the worldline of any individual  particle 
is denoted by
$ (D \bar{p} / Dt) $ resp. $ (D \bar{S} / Dt) $.
Hence,
\begin{eqnarray}
&&\mbox{freefalling-nonrotating particles:}
\nonumber
\\
&&\quad \, \, \frac{D \bar{p}}{Dt} \, = \, 0, \quad \quad 
  \frac{D \bar{S}}{Dt} \, = \, 0.
\end{eqnarray}

For arbitrary LONBs $ \bar{e}_{\hat{a}}, \,$  
the covariant derivative of vector fields  
follows directly from the definition of the Ricci connection coefficients
in
   Sect.~\ref{def.Ricci.coeff},
%
\begin{eqnarray}
(\nabla_{\gamma} \, \bar{V})^{\hat{a}} \, \, \, &=& \, \, 
\partial_{\gamma}  \,        V^{\hat{a}} \, + \,
(\omega^{\hat{a}}_{\, \, \, \hat{b}})_{\gamma} \, V^{\hat{b}},
\nonumber
\\ 
\nabla_{\gamma} \, \bar{e}_{\hat{b}} \, \, \, &=& \, \, 
\bar{e}_{\hat{a}} \, (\omega^{\hat{a}}_{\, \, \, \hat{b}})_{\gamma}.
\end{eqnarray}
%

The fundamental observational rocks 
for general relativity
are 
 $(\vec{E}_{\rm g}, \, \vec{B}_{\rm g})$,
measured by the acceleration resp. precession 
of freefalling-nonrotating test-particles
resp. gyroscopes which are quasistatic relative to the observer, 
as in the experiments of Galilei resp. Foucault. 
Apart from a minus sign and a factor 2, 
these measured fields 
are identical with Ricci's LONB-connection 
for a displacement 
along the observer's worldline.

\newpage

\section{Newton-inertial  motion
versus 
\protect\\
\quad \quad General-Relativity-inertial}
\label{inertial.motion}
%

This paper is based on two pillars:
\begin{enumerate}
\item Einstein's revolutionary concept of {\it inertial motion} 
as freefalling-nonrotating,  which replaces the first law of Newton,
his law of inertia,
\item The {\it Newtonian law}  on 
{\it relative acceleration} 
of neighbouring freefalling nonrelativistic test particles, 
spherically averaged.
\end{enumerate}
We never assume a 
 ``Newtonian limit of general relativity'',
and we never assume  ``weak gravity''.
%

From the above  two pillars, 
it immediately follows that space-time is {\it curved}, i.e.
the curvature of $ \, [t, \, x^i] \, $-plaquettes  does not vanish,
a straighforward, but revolutionary insight demonstrated in
    Sec.~\ref{revolutionary.concept}.

Incorporating the local Minkowski metric of special relativity 
leads  directly to Riemannian geometry.

From these inputs alone, 
\begin{center}

      We derive the {\it exact} curvature component 
      $ \, R^{\hat{0}  \hat{0}} \, (P)$ \\
of general relativity
\\
      from Newtonian relative acceleration experiments
\\ 
      of freefalling test-particles,
%
\begin{eqnarray}
[R^{\hat{0}  \hat{0}}]_{\rm GR \, exact}    &=&    
- \, [\frac{\partial}{\partial r}
<a^{\rm relative}_{\rm radial}>^{\rm quasistatic}_{\rm spherical \, av.}]_{r = 0}.
\nonumber                                     
\end{eqnarray}
     %
\end{center}
where one takes the spherical average of the relative radial acceleration of
freefalling quasistatic test-particles starting at radial distance  $r$.

Einstein's Ricci-0-0 curvature of general relativity is
{\it exactly} determined by an experiment with
{\it quasistatic} test-particles.

%

\subsection{Newton's method to find inertial motion}
\label{Newton.method.inertial.motion}

Newton's first law, the law of inertia,
states that  force-free particles 
move in  straight lines and  at constant velocities.  
The same law of inertia holds in special relativity.

Newton's law of inertia 
is valid only relative to {\it inertial reference frames.}
Inertial frames are defined as
those frames in which the law of inertia is valid.
Inertial frames can be constructed operationally 
either using 
three force-free
particles  moving in non-coplanar directions,
or using one force-free particle
plus two gyroscopes with non-aligned spins axes.

%
\begin{itemize}
\item There is a {\it grave problem} with this standard formulation 
      of Newton's first law: 
{\it force-free particles do not exist} e.g. 
for  {\it solar-system dynamics},
gravitational forces are always present,
therefore the operational construction of  
an inertial frame 
is a very difficult task. 
\end{itemize}
%
In a few cases, gravity is irrelevant:
(1)~for particle physics, 
gravity is negligible, 
since the relative magnitude of  the gravitostatic  
compared to  electrostatic force 
between two protons is extremely small, $\, 10^{- 36}$,   
(2)~for the motion of polished balls on a polished horizontal table, 
gravity cannot act,
it is orthogonal to the table.

Newton's  solution for this problem:
He
wrote in the {\it Principia} 
      \cite{Principia}
that for finding ``true motion'' equivalent to  ``absolute motion''
(i.e. motion relative to inertial frames),
distinguished from ``relative motion'', 
and in particular {\it true (absolute) acceleration}
(i.e. acceleration relative to inertial frames),
distinguished from relative acceleration,
one must first {\it subtract}  the {\it gravitational effects} 
of all celestial bodies. 
In the {\it Scholium} on space and time,
at the end of the initial ``Definitions'', 
Newton writes 
on p.~412 of \cite{Principia}:
``True motion is neither generated nor changed except by {\it forces}
impressed upon the moving body.''
On p.~414: 
\begin{itemize}
\item
``It is certainly very difficult to find out the true motions of individual
bodies.
... Nevertheless, the case is not hopeless. For it is possible to draw
evidence ... from the {\it forces} that are the {\it causes} ... 
of the {\it true motions.''}
\end{itemize}
In order to find inertial frames, Newton invokes the forces from 
his second law. Newton implies that without his second law,
his first law would be ``hopeless''.~---
On p.~415, at the very end of this Scholium, Newton concludes:
``In what follows, a fuller explanation will be given of  
how to 
{\it determine true motion from their causes} ... 
{\it For this was the purpose for which I composed the
following treatise}.''

Essentially by this method, 
the local inertial center-of-mass frame for the solar system 
is determined today, the ``International Celestial Reference Frame'' 
of 1998.

Note the dramatic difference between the two aspects of inertial frames:
(1) It is easy to establish
the {\it nonrotating frame} by using two gyroscopes (Foucault) 
or, much less precisely, with the rotating-bucket experiment of Newton.
(2) It is very work-intensive to establish 
a {\it frame without linear acceleration} relative to an inertial frame
in the mechanics of the solar system.

It is remarkable that most  
textbooks on classical mechanics
pretend that Newtonian inertial frames  
can be constructed operationally
using force-free particles,
although force-free particles 
do not exist for the mechanics of the solar system.

\subsection{Poincar\'e's man under permanent cloud cover}

An entirely different point was discussed by  Henri Poincar\'e in 
{\it ``Science and Hypothesis''}  in 1904 
   \cite{Poincare.inertial.frames}:
``Suppose a man were transported to a planet, the sky of which was
{\it constantly covered} with a thick curtain of {\it clouds}, so that he
could never see the other stars. 
On that planet he would live as if it were
isolated in space. 
But he would notice that it rotates, 
either by measuring the planet's ellipsoidal shape 
(... which could be done by purely geodesic means), or by repeating the
experiment of Foucault's pendulum. The absolute rotation of this planet might
be clearly shown in this way.''

On such a planet with permanent dense cloud cover,
it is impossible to use Newton's method
``to draw
evidence ... from the {\it forces} that are the {\it causes} ... 
of the true motions,'' since the Sun is invisible.
On such a planet,
it is {\it impossible to decide,}
whether the planet is on an orbit around the Sun, 
i.e. in {\it non-inertial motion} in the sense of Newton, 
or moving in a straight line with constant velocity,
i.e. in {\it inertial motion} in the sense of Newton.
If the planet was sufficiently small, tidal effects from the Sun
would be undetectable.
It follows that
\begin{itemize}
\item 
{\it Newton's first law} is 
{\it operationally empty}  
for a person on a very small planet with permanent cloud cover: 
{\it One pillar} of Newtonian physics is {\it destroyed}. 
\end{itemize}

\subsection{Einstein's revolutionary concept of 
\protect\\
inertial motion}
\label{revolutionary.concept}

But one cannot remove one pillar from a theory 
without putting in its place a new pillar:

\begin{itemize}
\item The new pillar is 
Einstein's {\it revolutionary re-definition} of {\it inertial motion}: 
%
Inertial motion and the local inertial frame
in general relativity
are operationally defined
by a freefalling particle together with
the spin axes (not parallel) of two comoving gyroscopes, 
i.e. {\it freefalling-nonrotating motion},
\end{itemize}
%

Freefalling particles
      determine and define 
      the {\it straightest timelike worldlines, geodesics.} They have    
      maximal proper time between 
      any two nearby points on the worldline, 
      $\int d\tau = \mbox{maximum}.$
%

%

%
%
%

Spacetime curvature, 
i.e. curvature of  space-time plaquettes $[t, x]$ etc.
in distinction to space-space plaquettes $[x, y]$ etc.,
follows directly from
{\it Newtonian experiments}  on 
{\it relative acceleration} 
of neighbouring freefalling test particles
combined with Einstein's concept of inertial motion and geodesics:
In a Gedanken-experiment, 
consider a vertical well 
   drilled down through an ideal ellipsoidal Earth 
from the North Pole 
   through the center of the Earth 
to the South Pole. 
Drop a pebble and  afterwards another pebble.
The pebbles will fall freely (geodesic worldlines) from the North Pole 
to the center of the Earth, 
rise to the surface of the Earth at the South Pole and fall back again.
The {\it two geodesics} will {\it cross again and again,}
direct evidence of {\it space-time curvature} 
from {\it Newtonian experiments}.
In 
   Sect.~\ref{sec.curvature.from.Newton},
%
\begin{itemize}
\item we compute the space-time {\it curvature} component
$R^{\hat{0}}_{\, \, \, \hat{0}}
= ({\cal{R}}^{\hat{0}}_{\, \, \, \hat{i}})_{\hat{0} \hat{i}}$
of {\it general relativity} 
for $[0i]$ space-time plaquettes
{\it exactly} from 
{\it Newtonian experiments}  on 
relative acceleration 
of neighbouring freefalling test particles, 
quasistatic relative to $\bar{e}_{\hat{0}},$
spherically averaged, 
using the revolutionary concept of inertial motion 
as freefalling-nonrotating.
\end{itemize}
%


\newpage

\section{Equivalence of
\protect\\ 
\quad fictitious  and gravitational forces
\protect\\ 
\quad in equations of motion}
\label{sec.equiv.th}

We do not trace the history of the equivalence principle 
in Einstein's writings beginning in  1907.

Starting with Einstein's theory of relativity of 1915,
we prove the {\it two equivalence theorems} 
with their {\it exact explicit equations} of motion,
\begin{enumerate}
       \item 
             for a {\it non-inertial observer}  
            the {\it equivalence} and  {\it exact equality}
            of {\it gravitational} forces and {\it fictitious} forces
            in the {\it equations of motion},
                   Sect.~\ref{equiv.th.non.inertial.obs},
       \item for an {\it inertial observer}  
            the {\it exact explicit  vanishing} of $\, (d/dt) (p_{\hat{i}})$
            and $\, (d/dt) (S_{\hat{i}})$ for inertial particles
            in   Sect.~\ref{equiv.th.inertial.obs}.
    \end{enumerate}
%
Our explicit equalities presented hold,
if and only if
one uses 
our {\it adapted spacetime slicing} and
our {\it adapted LONBs}  presented in 
   Sect.~\ref{sect.adaped.slicing.LONBs.2}.

The reader might want to 
first read our results in 
   Sects.~\ref{equiv.th.non.inertial.obs} and
          \ref{equiv.th.inertial.obs},
afterwards
our method in 
   Sect.~\ref{sect.adaped.slicing.LONBs.2},
and finally our discussion of fundamentals
in    Sect.~\ref{sect.fundamentals}.

\subsection{Fundamentals}
\label{sect.fundamentals}

Fictitious forces, e.g. centrifugal forces seen by a rotating observer
in his rotating reference frame,
have been important 
since Huygens, Newton, Leibniz, and Hooke.
Fictitious forces are also called inertial forces.~---
Because the centrifugal force vanishes 
on the worldline of a rotating observer,
this force will not play any role 
for the equivalence 
between fictitious and gravitational forces 
in the equations of motion.


Equivalence principle for $\vec{E}_{\rm g}$:
For an {\it observer accelerated} with $\vec{a}_{\rm obs}$
relative to freefall, 
who is inside  a windowless elevator or windowless space-ship or in fog, 
and measures
the acceleration $\vec{a}_{\rm ff \, particle}^{\, (\rm relat.to \, obs)}$ 
of  freefalling test-particles at his position,
it is {\it impossible to  distinguish} between 
(1)~particle-acceleration caused 
by the {\it fictitious force} 
due to the observer's own acceleration relative to freefall,
and (2)~particle-acceleration
caused by  {\it gravitational forces}
due to the attraction by matter-sources (Earth, Sun, Moon, etc).
Therefore, 
this fictitious force and 
this gravitational force are {\it equivalent} in the {\it equations of motion}.

Equivalence principle for 
$\vec{B}_{\rm g}$:  
For an  {\it observer rotating} with angular velocity $\vec{\Omega}_{\rm obs}$
relative to comoving gyro-spin axes at his position,
who is in a windowless elevator or spaceship or in fog,
it is {\it impossible to distinguish}
(1)~a {\it fictitious Coriolis force} 
   $\, \vec{F} = 2 \, m \, [\vec{v} \times \vec{\Omega}_{\rm obs}] \,$
due to his own rotation,
from (2)~a {\it gravito-magnetic Lorentz-force}  
   $\, \vec{F} = m [\vec{v} \times \vec{B}_{\rm g}] \, $
from the gravitational field $\vec{B}_{\rm g}$ 
generated by mass-currents.
The gravitomagnetic field $\vec{B}_{\rm g}$ has been 
postulated by Heaviside in 1893 
   \cite{Heaviside}
in analogy to the magnetic field in electromagnetism,
and for general relativity, the gravito-magnetic field is defined  
in Eqs.~(\ref{def.B} - \ref{Coriolis.force}).
Both forces are  
   written here for
   nonrelativistic particles.
For an observer in a windowless spaceship, 
a fictitious Coriolis force
and a gravito-magnetic Lorentz-force are fundamentally indistinguishable,
hence equivalent.



For an observer in a windowless spaceship, 
the {\it mass-sources} of gravitational fields (Sun, etc) are {\it unknown}.
Therefore, for such an observer 
the {\it gravitational field equations} are {\it useless}, 
in 19th-century physics Gauss's law for gravity,
$\, \mbox{div} \, \vec{E}_{\rm g} = - 4\pi G_{\rm N} \, \rho_{\rm mass}.$

Newton has emphasized that 
two reference systems rotating relative to each other 
are {\it equivalent, unless} one considers the 
forces, 
   Sec.~\ref{Newton.method.inertial.motion}. 
Einstein's principle of {\it equivalence is no longer valid}, 
as soon as 
(1) the source-masses of gravity are seen 
(observer not ``under a permanent cloud cover'') and  
(2) the {\it gravitational forces} are explicitely {\it known} by solving
the gravitational field equation of Gauss-Newton  or Einstein.
Conclusion: the Principle of Equivalence is {\it not valid} 
at the level of the gravitational
{\it field equations}.

%
\begin{itemize}
\item 
         The {\it equivalence principle} states 
         that {\it fictitious forces} 
         are equivalent to {\it gravitational forces}
         in the {\it equations of motion}.
         The equivalence principle 
         holds  for {\it first time-derivatives} 
         of particle momenta  and spins 
         {\it on} the {\it worldline} of the primary observer. 
\item    But the equivalence {\it cannot  hold} for
         {\it relative acceleration (tidal acceleration)} and for 
         {\it relative precession} 
         of two particles at different positions, 
         i.e. the equivalence cannot hold 
         for {\it second derivatives} 
         of particle momenta and spins,
         one derivative in time  
         and one derivative in space 
         (to compare particles at different locations):
         The equivalence cannot hold for
         the {\it curvature} tensors.~---
         For an accelerated observer in Minkowski spacetime, 
         there are no tidal forces, 
         no relative acceleration of freefalling particles, 
         and no spacetime-curvature.
\item
       The equivalence {\it cannot hold}, 
       if in the equations of motion at $P$,
       the particle is  initially, at $t_P$, 
       {\it off} the {\it primary observer's world-line},
       because the {\it tidal forces} would be relevant.
%
\item  The equivalence {\it cannot hold} in the 
       gravitational {\it field equations}
       of Einstein or Gauss,
       $\mbox{div} \vec{E}_{\rm g} 
        = - 4 \pi \, G_N \, \rho_{\rm mass},$
       with their {\it matter-sources}.~--- 
       For an accelerated observer in
       Minkowski space, there are no matter sources.
\end{itemize}
%

We focus on {\it one primary observer}.
In the equations of motion,
e.g. in $\, [(d/dt) \, (p_{\hat{i}})](t_1) 
= \lim_{\delta t \rightarrow 0} 
[p_{\hat{i}} (t_1 + \delta t) - p_{\hat{i}} (t_1)]/\delta t$, 
spacetime {\it curvature}  is {\it absent,} 
if and only if {\it particles start} at $t_1$
{\it on} the {\it worldline} of the {\it  primary observer}.

Many {\it auxiliary local observers} are needed 
to measure the momenta $\vec{p}$ and spins $\vec{S}$ of  test-particles at $t
= t_1 + \delta t,$
when  they are no longer on the worldline of the primary observer.
We shall see that 
our auxiliary observers cannot  be inertial, even if the primary observer is
inertial
(unless spacetime curvature vanishes).
Every auxiliary local observer, 
with his worldline through $Q$ and $\bar{u}_{\rm obs} (Q) = \bar{e}_{\hat{0}}
(Q)$ and with his local spatial axes $\bar{e}_{\hat{i}}$,
is in a one-to-one relationship with
the Ricci connection coefficients 
$\, (\omega^{\hat{a}}_{\, \, \, \hat{b}})_{\hat{0}} \,$ at $ Q.$


An observer can be  non-inertial in exactly {\it  two fundamental}  ways:
\begin{enumerate}
\item
observer's {\it acceleration} relative to freefall,
\item
observer's {\it rotation} relative to spin axes of comoving
gyroscopes.
\end{enumerate}
Correspondingly, there are exactly {\it two fictitious forces} 
in the equations of motion on the worldline of a non-inertial observer,
equivalent to {\it two gravitational forces}:
\begin{enumerate}
\item
the fictitious force measured by an {\it observer accelerated} relative
to freefall,
which is equivalent to Newton's force in a {\it gravito-electric} field, 
$\vec{g}~\equiv~\vec{E}_{\rm g}$,
\item
the fictitious {\it Coriolis force} measured by an {\it observer rotating}
relative to gyro axes,
which is  equivalent to 
the {\it gravito-magnetic} Lorentz force due to $\vec{B}_{\rm g}$ postulated by
Heaviside in 1893. 
\end{enumerate}
%
%

%
The remaining  two of the four fictitious forces 
   \cite{Landau.Lifshitz}
cannot contribute in the equations of motion
{\it on} the  worldline of a primary observer,
because these forces {\it vanish on his worldline}.
Therefore these {\it two fictitious forces cannot occur} 
in the {\it equivalence theorem},
%
\begin{enumerate}
\item
the fictitious {\it centrifugal force,} 
\begin{eqnarray}
\vec{F}/m 
\, &=& \,  [\, \vec{\Omega}_{\rm obs} 
\times [\, \vec{r}  \times \vec{\Omega}_{\rm
       obs} \, ] \, ],
\nonumber
\end{eqnarray}
%

\item
the fictitious force due to an {\it observer's non-uniform rotation},
\begin{eqnarray}
\vec{F}/m \, &=& \,  
[ \, \vec{r} \times (\, d\vec{\Omega}_{\rm obs}/dt \, ) \, ],
\nonumber
\end{eqnarray}
\begin{center}
{\it both vanish} on the worldline\\ of the primary rotating observer, $r = 0.$
\end{center}   
\end{enumerate}
%



To put this subsection ``Fundamentals'' in perspective,
we print 
the criticism of the equivalence principle by  J.L.~Synge
   \cite{Synge}:
 ``Does the principle of equivalence mean that 
the effects of the gravitational field are indistinguishable from the 
effects of the observer's acceleration? If so, it is false. ...
Either there is a gravitational field or there is none, according to the
Riemann tensor. ... This has nothing to do with the observer's worldline.
 ... The Principle of equivalence ... should now be buried.''

Synge mixes up two entirely different levels, 
first time-derivatives of momenta
versus second derivatives of momenta around a plaquette:
\begin{enumerate}
\item  
{\it Equations of motion, acceleration}
with {\it first time-derivatives} of momenta 
for particles starting  at one point on
the observer's worldline.~---
The {\it equivalence principle} holds for the equations of motion,
where curvature is invisible.
The equivalence principle 
makes {\it no statements} about  relative acceleration and
curvature.

\item 
{\it Equations of relative acceleration} of neighbouring
freefalling test-particles
{\it starting at two different positions} with  
{\it second derivatives} of momenta around a plaquette 
with one time-derivative
and one spatial derivative 
to obtain relative acceleration, 
hence space-time {\it curvature}.
\end{enumerate}
%

\subsection{The primary-observer-adapted 
\protect\\
space-time slicing and LONB-field}
\label{sect.adaped.slicing.LONBs.2}

For the theorem of equivalence 
between fictitious forces and gravitational forces, 
the equations of motion 
$\, \nabla_{\bar{p}}~\bar{p}~=~0 \,$ and  $\, \nabla_{\bar{p}}~\bar{S}~=~0 \,$
for inertial particles
are not instructive by themselves,
because they do not make explicit 
the fundamental distinction  
between
inertial and non-inertial observers.

The more explicit equations of motion for inertial particles,
the geodesic equation,  
$\, (d/dt) \, p_{\hat{i}} \, + \, (\omega_{\hat{i} \hat{a}})_{\mu} \, p^{\hat{a}} \,
(dx^{\mu} / dt) = 0, \, $
and the Fermi spin-transport equation,
$\, (d/dt) \, S_{\hat{i}} \, + \, (\omega_{\hat{i} \hat{a}})_{\mu} \, S^{\hat{a}} \, 
(dx^{\mu} / dt) = 0, \, $
are not instructive by themselves,
because they contain, depending on the formalism used,
%
$6 \times 4 = 24$ \, 
Ricci LONB-connection coefficients  
$ \, (\omega_{\hat{a}   \hat{b}})_{\hat{c}}, \,$ resp.
Cartan LONB-connection coefficients
$ \, (\omega_{\hat{a}  \hat{b}})_{\gamma},$
since they are antisymmetric in the lower-index pair $[\hat{a} \hat{b}],$
hence 6 pairs, and they have 4 displacement indices.~---
There are $4 \times 10 = 40$ 
Christoffel connection coefficients 
$(\Gamma^{\alpha}_{\, \,  \beta})_{\gamma},$
since they are symmetric in the last index-pair $(\beta \gamma)$,
hence 10 pairs, and there are 4 values for $\alpha.$~---
%
These 40 resp. 24 connection coefficients are highly uninstructive.


For the exact equations of motion demonstrating explicitely 
the equivalence theorems 
for both {\it non-inertial} 
for  {\it inertial observers}, 
we need a 
{\it primary-obserer-adapted spacetime splitting} 
and {\it LONB-field:}
\begin{itemize}
   \item our primary-observer-adapted space-time splitting uses 
         {\it fixed-time slices} which are generated by
         {\it radial 4-geodesics} starting Lorentz-orthogonal
         to the worldline of the primary  observer,
   \begin{center}
         \quad \quad  space-time slicing 
         by radial 4-geodesics \\
         from primary observer,
   \end{center}
   \item the time-coordinate $t$ all over the fixed-time 
         slice $\, \Sigma_t \,$
         is defined as the time measured 
         on the wristwatch of the primary observer, 
   \item our {\it Local Ortho-Normal Bases},  LONBs,   
         are {\it radially parallel}, i.e. 
         parallel along {\it radial 4-geodesics} 
         from  the primary observer's LONB,
   \begin{center}
         LONBs radially parallel \\
         to primary observer's LONB,
   \end{center}
   \item  3-coordinates on the slices $ \, \Sigma_t \,$ 
          are {\it not needed} 
          in the equations of motion.
\end{itemize}
%

In the equations of motion, for $\,  (d/dt) \, (p_{\hat{i}}), \, $ 
and for the equivalence theorems, 
we work with LONBs exclusively,
hence with Ricci connection coefficients
$\, (\omega^{\hat{a}}_{\, \, \, \hat{b}})_{\hat{d}}.   \, $
We do so, because the LONB-components $\, p_{\hat{i}} \, $ 
can be {\it measured directly},
while the {\it coordinate-basis components}
$\, p_i \, $
{\it cannot be measured without} 
having first {\it solved Einstein's equations} 
to obtain the metric in the coordinate basis, $\, g_{\mu \nu}.$

            Our slicing and  LONB-field
            are needed
            for the primary observer's direct determination 
            of momentum components $\, p_{\hat{i}} (t_1 + \delta t) \, $  
            and spin components  $\, S_{\hat{i}} (t_1 + \delta t)\, $        
            {\it off} his worldline:
The primary observer measuring  
$\, \vec{p} \, (t) \, $ of a particle off his worldline,
needs a {\it parallel transport} of the particle's $\, \bar{p} \, (t) \, $
on the radial geodesic on $\, \Sigma_t \,$ to his own LONB at the time $t$.
In this parallel transport, 
the {\it components} $\, p_{\hat{i}} $ stay {\it unchanged},
because we have chosen the LONBs radially parallel.
This makes the explicit equations simple.

Our crucial  result: 
with our    
{\it  LONBs radially parallel},
%
\begin{itemize}
\item
    the connection coefficients for 
    {\it spatial displacements}
    vanish {\it on} the entire worldline of the primary observer,  
%
\begin{eqnarray}
   [ \, (\omega_{\hat{a} \, \hat{b}})_{\hat{i}} \, ]_{\, (r \rightarrow 0, \, t)}  
    \, \, &=& \, \,  0, 
\label{all.connection.coeff.zero}
\\
    \hat{i}  = (1,2,3),  \, \, && \, \,    \hat{a},\hat{b} = (0,1,2,3).
\end{eqnarray}
%
\item For a {\it non-inertial observer}, 
      the connection coefficients for {\it displacements in time}
      are non-zero,
\begin{eqnarray}
     (\omega_{\hat{i} \, \hat{0}})_{\hat{0}}   
    \, \, &=& \, \,  E_{\hat{i}}^{(\rm g)}, 
\nonumber
\\ 
     (\omega_{\hat{i} \, \hat{j}})_{\hat{0}}   
    \, \, &=& \, \,  B_{\hat{i} \hat{j}}^{(\rm g)}. 
\end{eqnarray}
\item
 For an {\it inertial} primary observer, 
      {\it all} connection coefficients 
      {\it vanish on} the primary observer's entire worldline, 
\begin{eqnarray}
&&
[(\omega_{\hat{a} \hat{b}})_{\hat{d}}]_{(r \rightarrow 0, \, t)} \, \, = \,
\,  0,
\quad \, \, \,   \hat{a}, \hat{b}, \hat{d} = (0,1,2,3).
\label{equiv.theorem}
\end{eqnarray}
%
\end{itemize}
%

A non-inertial and an inertial observer have necessarily different worldlines
(displacement curves), but only the {\it tangent vector} in $P$ matters in the
connection-1-form coefficients $\, (\omega_{\hat{a} \hat{b}})_{\hat{c}}$.
The tangent-vector in $P$ is the same, if the two observers are instantaneously
comoving in $P$.

Our {\it radially parallel} spatial LONB-field  $\, \bar{e}_{\hat{i}} \,$ 
in curved spacetime
is 
{\it consistent} with the {\it 3-globally parallel} basis vector-field 
$\, \partial_i \,$ 
in Newton's Euclidean 3-space with Cartesian coordinates.

In the equations of motion, 
evaluated on the worldline of the observer,
it is irrelevant that our LONB's 
$\, (\bar{e}_{\hat{0}},  \,  \bar{e}_{\hat{x}}, 
 \,  \bar{e}_{\hat{y}},  \,  \bar{e}_{\hat{z}})$
are not parallel in the $\, (\theta, \, \phi)$ directions.

Our {\it radially parallel} $\, \bar{e}_{\hat{0}} $ field 
in curved spacetime
is unique in being consistent with  {\it Newtonian} observers at a given time being
at {\it relative rest}, 
hence basis vectors $\partial_t$ {\it spatially parallel} 
in the language of affine geometry of Felix Klein of 1872. 

Our extended fields of LONBs define an {\it extended reference frame}
for an inertial or for a non-inertial primary observer 
along their entire worldlines.~---
Spatially, this holds outwards until radial 4-geodesics cross, 
e.g. from multiply-imaged quasars.

\subsection{Equivalence theorem for inertial observers}
\label{equiv.th.inertial.obs}

The equivalence theorem for inertial observers states: 
Along the {\it entire worldline} 
of a primary {\it inertial observer} (freefalling-nonrotating),
there exists an
{\it adapted slicing} and {\it adapted LONBs},
which is our slicing and our LONBs given in 
     Sect.~\ref{sect.adaped.slicing.LONBs.2},    
such that the equations of motion for relativistic 
{\it inertial test-particles,}
the geodesic equation and Fermi transport equation,
   Eqs.~(\ref{geodesic.Fermi.spin.transport.eq}),
give zero on the right-hand-side,
they have the {\it same form}
as for free particles in special relativity with Minkowski coordinates,
%
\begin{eqnarray}
\mbox{inertial particles:} \quad  
\frac{d}{dt} \, (p_{\hat{i}})   &=&   0, \quad \, \,     
\frac{d}{dt} \, (S_{\hat{i}}) \,  =  \, 0, \quad
\label{equiv.princ}
\end{eqnarray}
where $t$ is the time measured on the worldline of the 
primary observer.

For an inertial observer and with our adapted slicing
and adapted LONBs,
{\it all} Ricci  
connection coefficients
$(\omega_{\hat{a} \hat{b}})_{\hat{c}}, \, $
vanish.
For an inertial observer with a general
non-adapted slicing and non-adapted LONBs,
there are  
   $6 \times 4 = 24$ non-vanishing Ricci 
connection coefficients
$(\omega_{\hat{a} \hat{b}})_{\hat{c}}, \, $
which would be needed 
in the geodesic equation and 
the Fermi spin-transport equation
with non-adapted LONBs 
and  non-adapted spacetime-slicing.


For an inertial observer
with our adapted slicing and LONBs,
the equivalence theorem of general relativity states
that
{\it  all gravitational fields} vanish
   which is equivalent with the vanishing of
{\it all 24 Ricci connection coefficients} 
$\, (\omega_{\hat{a} \hat{b}})_{\hat{c}}. \, $

%

\subsection{Equivalence theorem for noninertial observers}
\label{equiv.th.non.inertial.obs}

Our three fundamental results  
for the equivalence theorem in the equations of motion
of general relativity 
for non-inertial observers are given first, 
the derivations follow afterwards:
%
\begin{itemize}
\item
Our results are explicitely valid, if and only if one uses 
our {\it primary-observer-adapted slicing} of spacetime, 
generated by radial 4-geodesics starting Lorentz-orthogonal to 
the primary observer's worldline, and
our {\it primary-observer-adapted LONBs,}  which are radially parallel 
to the LONB of the primary observer
as described in Sec.~\ref{sect.adaped.slicing.LONBs.2}.  
\item
First result: 
For any  test-particle 
and an {\it observer}  
who is {\it non-relativistic} 
relative to this particle, 
our {\it exact} equations of motion in 
{\it arbitrarily strong gravitational fields}
of {\it general relativity}
are {\it identical}
with the 19th-century Newton-Heaviside-Maxwell  equations of motion,
\begin{eqnarray}
&&\quad\frac{d}{dt}  (p_{\hat{i}}) \, =
\label{Heaviside.eq.motion.3}
\\ 
&&\quad \,  \, \, \, \, \, 
m \,  [\vec{E}_{\rm g}  
             +  \vec{v} \times \vec{B}_{\rm  g}]_{\hat{i}}
 \quad \quad \mbox{gravity in general relativity},
\nonumber
\\
&&\quad+ \, \, \,
q \, \, [\vec{E}  
             +  \vec{v} \times \vec{B}]_{\hat{i}},
\, \,   \quad   \quad  \mbox{el.mag. in general relativity.}
\nonumber
\end{eqnarray}
Every term in this exact equation of {\it general relativity}
is {\it identical} to the terms known in the {\it 19th century}:
The conceptual framework is different in general relativity, 
but the equations are the same.

(1)~{\it Newton's gravitational force}, the {\it gravito-electric force} due to
$\, \vec{E}_{\rm g} = \vec{g}$ in Heaviside's notation of 1893,
defined for general relativity in Eq.~(\ref{def.E}),

(2)~{\it Heaviside's gravito-magnetic force} 
due to $\, \vec{B}_{\rm g},$ 
postulated 1893 
    \cite{Heaviside},
defined for general relativity in Eq.~(\ref{def.B}),
and
form-identical with the Lorentz magnetic force 
due to $\vec{B}$ of Amp\`ere and Maxwell,

(3)~{\it Maxwell's electromagnetic forces} due to $\, (\vec{E}, \, \vec{B})$. 
%
\end{itemize}
%

Our exact equation of motion of general relativity
in our
primary-observer-adapted spacetime slicing and LONBs,
   Eqs.~(\ref{Heaviside.eq.motion.3}), 
has only {\it two gravitational fields}, $\vec{E}_{\rm g}$ and $\vec{B}_{\rm
  g}$,
versus  {\it 40 Christoffel symbols,} utterly non-instructive, 
for general coordinates 
(both in general relativity 
and in Newtonian physics with its universal time).

%
\begin{itemize}
\item
Second result: For {\it relativistic} test-particles, 
the {\it exact} equations of motion of general relativity
in {\it arbitrarily strong} gravitational fields
and electromagnetic fields are,
%
%
\begin{eqnarray}
\frac{d}{dt}  (p_{\hat{i}}) &=& 
\varepsilon \,  [\vec{E}_{\rm g}  
             +  \vec{v} \times \vec{B}_{\rm  g}]_{\hat{i}}
\, \,    \mbox{gravity in GR}
\label{eqs.motion.p.noninertial.2}
\\
&+& 
q \, [\vec{E}  
             +  \vec{v} \times \vec{B}]_{\hat{i}},
\, \, \,  \,   \mbox{el.mag. in general relativity}
\nonumber 
\\
\frac{d}{dt} \, \varepsilon 
&=&  \varepsilon \, \, \vec{v} \cdot \vec{E}_{\rm g}
\quad \quad \quad \, \, \, \, \,  \mbox{gravity in general relativity}
\nonumber
\\
&+&  q \,  \, \vec{v} \cdot \vec{E},
\quad \quad \quad \, \, \, \, \, \mbox{el.mag. in general relativity.}
\nonumber
\end{eqnarray}
%
The energy and 3-momentum of the test-particle,
\begin{eqnarray}
\varepsilon &=& \mbox{total particle-energy} = \gamma \, m, \, \, \,
\gamma =  (1 - v^2/c^2)^{- 1/2},
\nonumber
\\
&& \mbox{energy of photon} =   \varepsilon_{\gamma} \, =    h \, \nu_{\gamma},
\nonumber
\\
\vec{p} &=& \mbox{3-momentum} = \gamma \, m \, \vec{v}, 
 \, \, \, \mbox{for photon}  \,   \, 
\vec{p} = h \, \nu_{\gamma} \, \vec{e}_{\hat{v}}.
\nonumber 
\end{eqnarray}
%
%
\end{itemize}
%


With our primary-observer-adapted space-time slicing and LONBs,
  \begin{enumerate}      
  \item
      the terms 
      due to  {\it electromagnetic} fields 
      in  general relativity
      are {\it explicitely identical}  with the equations 
      in  special relativity,
   \item 
      the terms 
      due to  {\it gravitational} fields in general relativity 
      are {\it form-identical} with the terms 
      for electromagnetic fields in Minkowski spacetime of special relativity,
      except for the replacements,
\begin{eqnarray} 
(\vec{E}, \,  \vec{B}) \, \, &\Rightarrow& \,    \,   
(\vec{E}_{\rm g}, \,       \vec{B}_{\rm g})\, 
\nonumber
\\
q \, \, \, &\Rightarrow& \, \, \, \varepsilon.
\nonumber
\end{eqnarray}
  \end{enumerate}
%

It is remarkable that these simple {\it exact} equations of motion 
for general relativity with {\it arbitrarily strong} gavitational fields
are missing in textbooks.

Eqs.~(\ref{Heaviside.eq.motion.3},  \ref{eqs.motion.p.noninertial.2}) 
are {\it exact} in {\it curved spacetime}. 
But equations of motion
have only {\it first derivatives} of particle momenta, 
therefore they are
{\it independent of curvature,} 
which needs second derivatives of vectors arond a spacetime plaquette. 



%
\begin{itemize}
\item Third result: Theorem of {\it equivalence} 
      between {\it fictitious forces}
      and     {\it gravitational forces} 
      in the exact equations of motion 
      of general relativity for relativistic particles
      in arbitrarily strong gravitational fields:

      (1) the equivalence between the fictitious force  
                   measured by an accelerated observer
          and a gravito-electric field $\vec{E}_{\rm g}$ 
                   generated by sources of matter, 

      (2) the equivalence between the fictitious Coriolis force
                   measured by a rotating observer
          and a gravito-magnetic Lorentz force from 
          a gravito-magnetic field~$\vec{B}_{\rm g}$
          generated by matter-currents.
\end{itemize}
%

Our fundamental definitions
of   $\vec{E}_{\rm g}$
and  $\vec{B}_{\rm g}$ in 
     Eqs.~(\ref{def.E}) 
      and (\ref{Omega.vs.B}), 
inserted in the exact equation of motion of general relativity, 
     Eq.~(\ref{eqs.motion.p.noninertial.2}),
for particles free of non-gravitational forces
give, 
\begin{eqnarray}
\frac{d}{dt} \, (p_{\hat{i}}) \, &=& \,  
\varepsilon \, \, [ \, - \, \vec{a}_{\, \rm obs} 
\, +  \, 2 \, \vec{v} \times \vec{\Omega}_{\, \rm obs} \, ]_{\hat{i}}
\nonumber
\\
&=& \, 
\varepsilon \, \,  [ \, \vec{E}_{\rm g} 
\, +  \, \vec{v} \times \vec{B}_{\rm g} \, ]_{\hat{i}}.
\label{equiv.th.final.eq}
\end{eqnarray}
%
These equations hold 
for relativistic particles in general relativity.~---
The first of these two equations agrees with Landau and Lifshitz,
{\it Classical Mechanics} \cite{Landau.Lifshitz},
   Eq.~(39.7)
for the nonrelativistic case.

      The {\it exact equivalence} 
      between {\it gravitational} forces and {\it fictitious forces} 
      in the {\it equation of motion} for general relativity
      and for relativistic particles
is evident
      by comparing the two right-hand-sides in
      Eq.~(\ref{equiv.th.final.eq}),
%
\begin{eqnarray}
&& \quad \quad \quad \quad \mbox{equivalence of}
\nonumber
\\
&&\mbox{fictitious and gravitational forces:}
\nonumber
\\
&& \quad \quad \quad  \, \, \,
\vec{a}_{\, \rm obs} \, \, \, \, = \, - \, \vec{E}_{\rm g},
\nonumber
\\
&& \quad \quad \quad  \, \, \,
\vec{\Omega}_{\, \rm obs} \, \, \, = \, \, \,  \frac{1}{2}\vec{B}_{\rm g}.
\label{equiv.eqs}
\end{eqnarray}

It is a crucial fact
that 
the {\it exact} 
   $\, \vec{E}_{\rm g} =  - \vec{a}_{\rm obs}, \, $
and
  $\,  \vec{B}_{\rm g}/2 =   \vec{\Omega}_{\rm obs},   \, $ 
can be {\it measured} with freefalling test-particles 
   and  gyroscopes
which are 
   {\it quasistatic} relative to the observer,
        Eqs.~(\ref{def.E},~\ref{Omega.vs.B}),
but they  {\it predict} the 
motion of {\it relativistic} particles
in
   Eq.~(\ref{equiv.th.final.eq}).

{\it Arbitrarily strong} 
gravito-electric and
gravito-magnetic fields
can be measured by {\it quasi-static} methods 
in the instataneous comoving inertial frame,
since only {\it infinitesimally small velocities} 
and
{\it gyro-precession angles}
arise after an infinitesimally short time.


Our results in this subsection
{\it look very familiar.}
But our results are entirely {\it new}\,:

(1) Our {\it exact} results for general relativity 
with  arbitrarily {\it strong gravitational fields} are new.
No ``Newtonian approximation'', 
no weak field limit, 
no perturbation theory.

(2) The exact definition of $\vec{E}_{\rm g}$    
in arbitrarily strong
gravitational fields is new, Sec.~\ref{grav.el.magn.fields}.

(3) The identity  
of $\, (\vec{E}_{\rm g}, \, \vec{B}_{\rm g}) \, $ 
with the Ricci connection for a displacement in time, 
$\, (\omega_{\hat{a} \hat{b}})_{\hat{0}}, \, $
is new,
Sec.~\ref{sect.E.B.Ricci.2}.

(4) Our exact equations of motion 
in arbitrarily strong gravitational fields,
   Eqs.~(\ref{Heaviside.eq.motion.3}, \ref{eqs.motion.p.noninertial.2}),
are new.

(5) Our equations 
   Eqs.~(\ref{equiv.th.final.eq}, \ref{equiv.eqs})  
for the equivalence of fictitious and gravitational forces
in the equations of motion are new.


It is remarkable that most textbooks on general relativity do not discuss 
the equivalence of
{\it fictitious} forces and {\it gravitational} forces
in the equations of motion.
The few textbooks, which do dicuss this equivalence, are not specific about
which fictitious force is equivalent to which gravitational force.
But most important: these textbooks do not give equations.
Words without equations is unususal for an 
important topic in theoretical physics.

It is also remarkable that in textbooks  
there is no discussion of measurements by rotating observers, 
  no equation demonstrating the {\it equivalence} 
  of the
  {\it fictitious Coriolis force} 
  with the
  {\it gravitomagnetic Lorentz force},
no discussion of the fundamental impossibility to 
distinguish these two forces in the {\it equations of motion}.

The {\it old paradigm} 
for {\it fundamentals} in general relativity
has been that one should work with general coordinates.
Only 
for applications to special situations, Schwarzschild, Kerr, 
Friedmann-Robertson-Walker, should adapted coordinates be used.

The {\it new paradigm} 
for {\it fundamentals} in general relativity
is that one must choose adapted spacetime slicing and adapted LONBs
to exhibit 
(1) the identity of
19th-century Newtonian equations of motion
and the equations of motion of general relativity 
for nonrelativistic test-particles,
(2) the identity of
the equations of motion of special relativity in an electromagnetic field
and the  equations of motion of general relativity 
with the obvious replacements 
$\, (\vec{E}, \,        \vec{B})    \Rightarrow  
    (\vec{E}_{\rm g}, \, \vec{B}_{\rm g}) \, $
and $\, q \Rightarrow \varepsilon.  \,$

The {\it derivation} of our {\it equations of motion} 
in gravitational plus electromagnetic fields,
    Eqs.~(\ref{eqs.motion.p.noninertial.2}),
follows from (1)~the geodesic equation
    plus the electromagnetic term and
(2) our primary-observer-adapted space-time slicing and
LONBs,
%
\begin{eqnarray}
\frac{d}{dt} \, p_{\hat{i}} 
\, &=& \, - \, (\omega_{\hat{i} \hat{a}})_{\mu} \, p^{\hat{a}}
\, (dx^{\mu}/dt)
\, + \,
q \, F_{\hat{i}\hat{a}} \, u^{\hat{a}}
\nonumber
\\ 
&=& - \, (\omega_{\hat{i} \hat{a}})_{\hat{0}} \, p^{\hat{a}}
\, + \,
q \, F_{\hat{i}\hat{a}} \, u^{\hat{a}}.
\label{geodesic.eq.3}
\end{eqnarray}
With our {\it primary-observer-adapted} slicing and LONBs,
      {\it radially parallel LONBs}, of 
Sec.~\ref{sect.adaped.slicing.LONBs.2}:
%
\begin{center}
connection coefficients
\\
nonzero only for displacement in time
\\ 
along worldline of primary observer,
\end{center}
\begin{eqnarray}
\frac{d}{dt} \, p_{\hat{i}} 
\, &=& \, \pm \, (\omega_{\hat{i}  \hat{a}})_{\hat{0}} \, p^{\hat{a}},
\label{our.geodesic.eq}
\\
 (\omega_{\hat{i} \hat{0}})_{\hat{0}} \, &=& \, - \, E^{(\rm g)}_{\hat{i}},
\nonumber
\\
 (\omega_{\hat{i} \hat{j}})_{\hat{0}} \, &=& \, - \,  \frac{1}{2} \, B^{(\rm
                                             g)}_{\hat{i} \hat{j}},
\nonumber
\end{eqnarray}
Using the 4-momentum components
$\, p^{\hat{0}} = \varepsilon, \, $ and
$\, p^{\hat{i}} = \varepsilon v^{i}, \,$
we obtain
   Eqs.~(\ref{eqs.motion.p.noninertial.2}). 

To derive  the equations of motion for a non-inertial observer,
    Eqs.~(\ref{eqs.motion.p.noninertial.2}),
the {\it Local Inertial Coordinate System} and the 
associated {\it Local Inertial Frame} 
used in all textbooks are {\it useless}. 

The  {\it time-evolution equation for spin,} 
the {\it Fermi spin transport equation,} 
   Eqs.~(\ref{geodesic.Fermi.spin.transport.eq})
simplify in our primary-observer-adapted LONBs and slicing.
Since the spin 4-vector in the rest frame of a particle 
has by definition no time-component, we have $\, p^{\hat{a}} S_{\hat{a}} = 0,
\,$
and 
we can eliminate $S_{\hat{0}}$  
\begin{eqnarray}
p^{\hat{a}} S_{\hat{a}} \, &=& \, 0 
\quad \quad \Rightarrow \quad \quad
S^{\hat{0}} \, = \, \vec{v} \cdot \vec{S},
\nonumber
\\
\frac{d}{dt} \, (S_{\hat{i}})
\, &=& \,
(\vec{S} \times \vec{B}_{\rm g}/2)_{\hat{i}}
\, \, + \, \,
(\vec{v} \cdot \vec{S}) \, \, E_{\hat{i}}^{(\rm g)},
\label{Fermi.transport.eqs.motion.p.noninertial.2}
\end{eqnarray}
where  $\vec{v} \equiv (d\vec{x}/dt)$ is the 3-velocity of the particle.
For a {\it comoving gyroscope,}
the second term in 
     Eq.~(\ref{Fermi.transport.eqs.motion.p.noninertial.2}) vanishes, 
and this equation for relativistic gyroscopes 
becomes identical to the equation for  nonrelativistic gyroscopes, 
    Eq.~(\ref{def.B}).
For a {\it non-comoving gyroscope,} the second term is the gravitational analogue of 
the  {\it Thomas precession} (1927) in classical electrodynamics 
\cite{Thomas.precession}.

\subsection{Connection-1-forms versus curvature tensor}

Textbooks give an {\it asymmetric status} to  
connections, 
e.g. 
Ricci's $(\omega^{\, \hat{a}}_{\, \, \, \, \hat{b}})_{\hat{c}}, $
versus curvature tensors, e.g. Riemann's
$(R^{\, \hat{a}}_{\, \, \, \hat{b}})_{\hat{c} \hat{d}}$.

For dynamics, the {\it equations of motion} 
and the {\it gravitational field equations}
are two equally important legs. 
In Wheeler's words: ``matter tells spacetime how to curve, curved
spacetime tells matter how to move.''

In the equations of motion, 
the {\it measured forces}  are encoded in the Ricci connection coefficients 
$(\omega^{\, \hat{a}}_{\, \, \,  \hat{b}})_{\hat{0}}$
with our adapted slicing and LONBs.
These observables are $\vec{E}_{\rm g}$ and  $\vec{B}_{\rm g}$, 
directly measured, 
the  {\it observational rock}, 
on which the equations of motion are built.~---
The equations of motion do not involve the Riemann tensor.

{\it Observers} are in a {\it one-to-one} relation with
Ricci's connection for a time-like displacement, 
$(\omega^{\, \hat{a}}_{\, \, \,  \hat{b}})_{\hat{0}}$.
Hence, observers and their worldlines are as fundamental 
as Ricci's connection for a time-like displacement.

\newpage

\section{Spacetime curvature 
from Newtonian experiments}
\label{sec.curvature.from.Newton}

In
     Sect.~\ref{revolutionary.concept}
we have considered
the Newtonian experiment  of dropping pebbles down a well drilled all the
way through the Earth.
We have explained, 
how this Newtonian experiment 
together with
Einstein's revolutionary concept of inertial motion and geodesics
gives direct evidence of space-time curvature.

We now show by explicit computation, 
how the 
spacetime curvature-coefficient $R^{\hat{0} \hat{0}}$ 
of general relativity
is
{\it completely determined}  
by  {\it Newtonian experiments}
alone.

In general relativity,
(3+1)-spacetime is {\it not embedded} in some higher dimensional spacetime,
and there is no observational evidence for such an embedding.
Therefore, curvature of (3+1)-spacetime always means {\it intrinsic curvature}.
Generally, intrinsic curvature is determined by measurements intrinsic to the
space considered.
The intrinsic curvature of a surface of an apple at a point $P$ 
can be determined by measuring angles and lengths on the surface of the apple near $P$.


\subsection{Curvature from deficit tangent-rotation angle
\protect\\   
around closed curves}
\label{deficit.rot.angle.tangents}

The {\it sum} of {\it inner angles} in a {\it triangle}
is always $180^0$ in the Euclidean plane.

For $N$-{\it polygons} 
as boundaries of  simply connected areas
in the Euclidean plane,
the sum of inner angles is not useful, it is $(N-2) \pi$ 
and goes to infinity
for $N \rightarrow \infty.$
It is much more useful to consider the 
sum of {\it tangent-rotation angles} $\alpha_i. \, $ 
For boundart-curves ${\cal{C}}$ 
of simply connected areas ${\cal{A}}$,
the integral of tangent-rotation angles 
is  always a {\it full
  rotation}, $\, 2 \pi, \,$
in the Euclidean plane,
%
\begin{eqnarray}
\sum_i \alpha_i^{(\rm tangents)}   &\Rightarrow&   
\oint_{\cal{C}} \, (\frac{d\alpha}{ds})^{(\rm tangents)} \, \, ds 
\, =  \, 2 \pi.
\nonumber
\end{eqnarray}

In a {\it non-Euclidean 2-space,} e.g. on the surface of an apple, 
$N$-polygons 
and   
curves ${\cal{C}}$ 
which are the boundary of a simply connected area ${\cal{A}}$,   
${\cal{C}} = \partial {\cal{A}}$,
have a sum, resp.  integral, of tangent-rotation angles
different from a full rotation of $2 \pi$,
\begin{eqnarray} 
\sum_i \alpha_i^{(\rm tangents)}   \, &\Rightarrow&   \, 
\oint_{{\cal{C}}} \,  (\frac{d\alpha}{ds})^{(\rm tangents)} \, \, ds
\, \,  \equiv \, \, 
2 \pi \, - \, \delta_{\cal{C}}, 
\nonumber
\\
\delta_{\cal{C}}
\, &\equiv& \, \mbox{deficit rotation angle}.
\label{deficit.rotation.angle.2} 
\end{eqnarray}

As an example, we compute the deficit angle $\delta$ for a
geodesic triangle on the surface of a spherical Earth:

(1) start at the North-pole and 
go South along the zero-degree meridian to the equator,

(2) go East along the equator to the $90^0$ meridian, 

(3) follow the $90^0$ meridian back to the North-pole. 
\\
This is a geodesic triangle with three tangent-rotation angles of $\pi/2$,
the total tangent-rotation angle is $3\pi/2$,
hence the deficit angle  is $\delta = \pi/2.$

The {\it intrinsic Gauss curvature} $R_{\rm G}$ at any point $P$  
of any 2-dimensional space can be operationally  defined by
      the {\it deficit angle} $\, \delta \, $ 
         divided by the measured {\it area} $\, A \, $ inside the curve
         $\, {\cal{C}} \, $ for the length of the curve around $P$ going to zero, 
\begin{eqnarray}
R_{\rm G} \, &=& \, \lim_{{\cal{C}} \rightarrow 0} \, 
( \, A^{-1} \, \delta_{{\cal{C}}} \, ). 
\label{Gauss.curvature.2}
\end{eqnarray}
%

For our chosen geodesic triangle on the surface of a spherical Earth,
the area of ${\cal{A}}$ can be measured entirely {\it on} the {\it surface} of the
Earth, an {\it intrinsic} measurement on the surface.
The result is
$A = r^2_{\rm Earth} \, (\pi/2)$.
The deficit angle is $\delta = \pi/2$, 
hence the Gauss intrinsic curvature is $R_{\rm G} = r^{-2}_{\rm Earth}.$

The primary definition of intrinsic curvature in elementary geometry
is given by 
   Eqs.~(\ref{deficit.rotation.angle.2}) and (\ref{Gauss.curvature.2}) 
in terms of the deficit rotation angle for geodesic triangles and
polygons.

\subsection{Curvature from LONB-rotation angle
\protect\\   
around closed curves}
\label{sect.Cartan.LONB.rot.closed.curve}

Cartan used a  different method
to  compute the same deficit angle $\delta_{\cal{C}}$ around a curve
${\cal{C}}.$

Cartan's method for computing curvature 
is missing in many textbooks on general relativity, 
and  it is not taught 
in most graduate programs in gravitation and cosmology.
Therefore, we give a  short introduction to Cartan's method.

Instead of using as a path 
an arbitrary closed curve ${\cal{C}} = \partial {\cal{A}}$,
and instead of considering the rotation of the tangent vectors along this
curve
${\cal{C}}$,
Cartan used 
(1)~the rotation of the chosen {\it LONB-field},
already crucial for 
    Secs.~\ref{grav.el.magn.fields} and \ref{sec.equiv.th}, 
(2)~the {\it closed path} along a {\it coordinate plaquette} 
$\, [ \, \bar{e}_{\alpha}, \, \bar{e}_{\beta} \,  ] 
\equiv [\, \partial_{\alpha}, \, \partial_{\beta}  \,]\, $
in the chosen coordinate system. 
Therefore, Cartan's method needs the LONB-rotation for a 
{\it displacement} in a {\it coordinate}, 
$\, (\omega_{\hat{i} \hat{j}})_{\mu}.$~---
Cartan needs a coordinatization, 
i.e. a mapping $\, P \Rightarrow x^{\mu},  \, $
but
the metric $\, g_{\mu \nu} \, $ is not needed to compute the Riemann tensor
with Cartan's method. 
The metric  $\, g_{\mu \nu} \, $ cannot be known until
Einstein's equations are solved  for the specific problem. 

In curved 2-space, LONB-rotations  are given by an angle, 
no need
for a rotation matrix,
$\, (\omega_{\hat{a} \hat{b}})_{\mu}  \Rightarrow \omega_{\mu}.$

In the {\it Euclidean plane}, 
the total LONB-rotation angle 
along any closed curve ${\cal{C}} = \partial {\cal{A}}$
is always zero,   
      \begin{eqnarray}
 (\frac{d\alpha}{dx^{\mu}})^{(\rm LONBs)} 
\, \equiv   \, \omega_{\mu}: \quad  \quad     \oint_{{\cal{C}}} 
      \, \omega_{\mu} \, \, dx^{\mu} \, &=& \, 0.
      \nonumber
      \end{eqnarray}
For a Cartesian LONB-field in Euclidean 2-space,  the LONB-rotation angle is
identically zero.
For the LONB-field aligned with polar coordinates in the Euclidean 2-space,
(1) the LONB rotation angle is nonzero 
for a displacement in the direction $\partial_{\phi}$, 
(2) the LONB-field is singular at the origin, 
hence it is not admitted, if the origin is inside ${\cal{A}}$.

In  {\it curved} 2-space, the total rotation angle  of LONBs
       relative to infinitesimal geodesic pieces 
       along the positively oriented boundary ${\cal{C}} = \partial {\cal{A}}$
       gives the {\it deficit angle,}
      \begin{eqnarray}
      \oint_{\cal{C}} 
      \, \omega_{\mu} \, \, dx^{\mu} \, &\equiv& \, - \,  \delta_{\cal{C}} \, \, \,
      \equiv \, \, - \, \, \mbox{deficit  angle}.
      \label{eq.deficit.angle.Cartan.3}
      \end{eqnarray}
The {\it line-integral} 
calls for 
the {\it 1-form} $\, \tilde{\omega} \,$
with  {\it covariant}  components $\, \omega_{\mu}$.
Using 1-form components $\,\omega_{\mu}  \, $
causes the {\it absence} of {\it metric factors} $\, g_{\mu \nu}  \,$ 
in the line-integral (\ref{deficit.angle.Cartan.2})  over curved space.      

For two areas touching along a piece of common boundary,
the total deficit angle is {\it additive},
because the LONB-rotation angles cancel
along the common boundary 
due to the opposite direction of the boundary curves.
The areas are also additive.
Therefore, the total deficit angle divided by the total area,
$(\delta_{\cal{C}} / A)$ is the same for any (simply connected) area
on a perfect sphere.~--- 
For the surface of an apple, 
an infinitesimal curve $\cal{C}$ around any point $P$ gives the local measure
of {\it intrinsic curvature}, the {\it Gauss curvature} at $P$ of 
   Eq.~(\ref{Gauss.curvature.2}),
\begin{eqnarray}
R_{\rm G} \, &=& \, \lim_{{\cal{C}} \rightarrow 0} \, 
( \, A^{-1} \, \delta_{{\cal{C}}} \, ). 
\nonumber
\end{eqnarray}
%

As an example of Cartan's method of LONB-rotaton angles
around a closed curve, 
we compute the intrinsic curvature 
of the surface of a spherical Earth 
We choose the LONB field 
$\, (\vec{e}_{\rm East}, \, \vec{e}_{\rm North} ).  \,$ 
For the closed boundary-path $\, \partial{\cal{A}},  \, $
we must {\it avoid} the {\it singular points} of this 
{\it LONB field,} 
North pole and South pole:

       (1) go South along $0^0$ meridian, start  an 
          infinitesimal  $\delta \theta$ 
         away from North pole, 
         end at equator: LONB rotation angle = $0,$ 

      (2) go East along equator to 90$^0$ meridian: 
         LONB rotation angle $= 0,$
 
      (3) go North along 90$^0$ meridian 
         to $\delta \theta$ before North pole:  
         LONB rotation angle $=0,$

      (4) go West along $\delta \theta =$ infinitesimal constant, 
         from 90$^0$ meridian
         to $0^0$ meridian: 

The total LONB-rotation angle is negative, $\alpha =  - \, \pi/2, \, \,$
      the deficit angle is positive, 
      $\,  \delta = \, \pi / 2, \, \,$ 
measured area 
      $ A =  (\pi/2) \, r_{\rm Earth}^2, \, \,$ 
intrinsic curvature $  \, R_{\rm G} \, = \, r_{\rm Earth}^{-2}$.

\subsection{Cartan's curvature equation   
in two dimensions}

We convert the line-integral for the deficit angle $\, \delta  \,$ of 
  Eq.~(\ref{eq.deficit.angle.Cartan.3})
using 
{\it Stoke's theorem,}
\begin{eqnarray}
\delta_{\cal{C}} 
\, &=& - \, 
\oint_{{\cal{C}} = \partial {\cal{A}}} \omega_{\mu} \, dx^{\mu} 
\, \, = \, - \, \int_{\cal{A}} \, (\mbox{curl} \, \vec{\omega}) \cdot 
\vec{dA}. 
\label{Cartan.curl.eq}
\end{eqnarray}
In two dimensions,
$\, \mbox{curl} \, \vec{\omega} \, $ 
and the Gauss curvature $\, R_{\rm G} \, $
are both {\it scalars} under rotation 
and space reflections (even under parity).
The rotation deficit angle $\, \delta \, $ and 
the oriented area $\, A \, $ are both {\it pseudoscalars}. 

%
\begin{eqnarray}
\mbox{Cartan curvature eq.}&&\mbox{in 2 dimensions}
\nonumber
\\
R_{\rm G} \quad \, \,  \, \, &=&  \quad \, \, - \, \mbox{curl} \, \vec{\omega}, 
\label{Cartan.curvature.eq.2}
\end{eqnarray}

We now need 
Cartan's tools: {\it connection 1-form}, {\it exterior derivative,}
and Cartan's {\it curvature 2-form}.

In curvilinear coordinates (in curved space or flat space), 
the {\it antisymmetric covariant derivative} 
of a vector field $\, \bar{V} \, $ in {\it covariant components}
is  
equal to the 
{\it ordinary antisymmetric derivative},
\begin{eqnarray}
[\, \nabla_{\mu} V_{\nu} - \nabla_{\nu} V_{\mu} \, ] \, &=&
\, [\, \partial_{\mu} V_{\nu} - \partial_{\nu} V_{\mu} \, ].
\nonumber
\end{eqnarray}
%
This holds, because the Christoffel connection coefficients 
$(\Gamma^{\alpha}_{\, \, \, \beta})_{\delta}$ 
are symmetric in the last two indices.

We shall see that
covariant derivatives  play no role 
      in Cartan's method of computing curvature.

In the calculus of forms (here for Riemannian geometry, which has a metric), 
a vector, if and only if it is given by its {\it covariant components} 
is called a {\it 1-form}
and denoted by a {\it tilde} (instead of an arrow or a bar), 
$\, \tilde{\sigma}$
with 1-form components $\, \sigma_{\mu}.$ 
The index structure of differential forms is trivial, therefore
one often drops indices,
\begin{eqnarray}
\mbox{1-form} \, \, \, \tilde{\sigma}
\, \, &\Leftrightarrow& \, \,
\mbox{1-form components} \, \, \, \sigma_{\mu}, 
\nonumber
\end{eqnarray}

The antisymmetric derivative of a 1-form $\tilde{\sigma}$ is 
called {\it exterior derivative}, 
denoted by the symbol $d$ in $\, d \, \tilde{\sigma}$,
\begin{eqnarray}
&&\quad \quad \, \,   \mbox{exterior derivative:} 
\nonumber
\\
&&d \, \tilde{\sigma} \, \,   \Leftrightarrow \, \,  
(d\, \tilde{\sigma})_{\mu \nu}  
\, \equiv \,   
\partial_{\mu} \, \sigma_{\nu}  - 
\partial_{\nu} \, \sigma_{\mu}.
\nonumber
\end{eqnarray}
The exterior derivative of a 1-form $\tilde{\sigma}$ produces
$ d \, \tilde{\sigma}$, a {\it covariant antisymmetric 2-tensor}, which
by definition is a {\it 2-form}.~---
General 2-forms are defined as antisymmetric covariant 2-tensors.

In 3-space, the {\it Hodge-dual star-operation} 
of $\, d \, \tilde{\sigma} \, $ 
is formed with the {\it Levi-Civita} tensor,
which is  defined by 
(1) totally antisymmetric tensor,
(2) $\varepsilon_{\hat{1} \hat{2} \hat{3}} \equiv + 1$
in a LONB with positive orientation,
as discussed in
   Eq.~(\ref{Levi.Civita}).
Applying the Hodge-dual star-operation 
to the exterior derivative of a 1-form in 3-space, 
$\, * \, d \, \tilde{\sigma}, \,$ gives the {\it curl}
of the vector field $\vec{\sigma}$, 
\begin{eqnarray}
(* \, d \, \tilde{\sigma})_{\lambda} 
\, &=& \,  
(\mbox{curl} \,  \vec{\sigma})_{\lambda}.
\nonumber
\end{eqnarray}

Hence, Cartan's curl equation for curvature in two spatial dimensions,
   Eq.~(\ref{Cartan.curvature.eq.2}), 
can be re-written in the calculus of forms, 
\begin{center}
Cartan's curvature scalar in 2 dim.:
\end{center}
\begin{eqnarray}
R \, &=&  \, * \, d \, \tilde{\omega} 
\, \, =  \, 
\frac{1}{2} \, \varepsilon^{\mu \nu} \,
[\, \partial_{\mu} \, \omega_{\nu} - \partial_{\nu} \, \omega_{\nu} \,].
\label{Cartan.2d.forms}
\end{eqnarray}

Cartan's equation for the curvature 2-form components in 2 dimensions,
\begin{eqnarray}
R_{\mu \nu} \, &=& \, 
\partial_{\mu} \, \omega_{\nu} - \partial_{\nu} \, \omega_{\mu}.
\label{Cartan.2form.comp.2dim}
\end{eqnarray}
%

The geometric meaning of the {\it two antisymmetric 
covariant indices} in $\, (d \, \omega)_{\mu \nu} \, $:

(1) find the infinitesimal Lorentz transformation of LONBs
under a {\it first displacement} along an observer's worldline,
$\, \bar{e}_{\nu} \equiv \partial_{\nu}, \, $ to obtain $\, \omega_{\nu}$,

(2) compare this quantity on a neighbouring worldline  separated by
a {\it second displacement} 
$\, \partial_{\mu} \, $ to obtain $\, \partial_{\mu} \, \omega_{\nu},$

(3) take the difference with the same  
in  {\it opposite order} $\, \mu \Leftrightarrow \nu \,$
to obtain 
$\, (\partial_{\mu} \, \omega_{\nu} - \partial_{\nu} \, \omega_{\mu}),$
which means going {\it around} the {\it closed displacement curve}
${\cal{C}} = \partial {\cal{A}}$,
the {\it coordinate plaquette}
$\, [\partial_{\mu}, \, \partial_{\nu}].  \,$

The displacement indices must necessarily be an 
{\it antisymmetric covariant index-pair}.
Only in this case is the displacement plaquette {\it closed}.
Example on Earth: if you go East by $10 ^o,$ go North by $10^o$,
go West by $10^o$, go South by $10^o$, you have a {\it closed plaquette}.
If instead, generated by $\partial_{\hat{a}}$, 
you go East 100 km, North 100 km, go West 100 km, go South
100 km, you have {\it no closed path}.
A coordinate-displacement plaquette, 
formed by $\, [\partial_{\mu}, \partial_{\nu}], \,$ 
is {\it closed}, as it must be for 
   Eq.~(\ref{eq.deficit.angle.Cartan.3}).
But a plaquette formed by $[\partial_{\hat{\mu}}, \partial_{\hat{\nu}}]$ 
is {\it not closed}.
This is the reason, 
why the displacement-plaquette indices  
must be coordinate indices.


The curvature 2-form $\, R_{\mu \nu} \, $ in 2 dimensions,
  Eq.~(\ref{Cartan.2form.comp.2dim}),
written suppressing  the  displacement index-pair, 
i.e. without the plaquette index-pair $[\mu \nu]$,
is denoted by script-${\cal{R}}$, 
\begin{eqnarray}
&&\mbox{curvature from deficit rotation angle in 2-dim,}
\nonumber
\\
&&\quad \quad \, \, \, \mbox{without LONB-rotation indices:}
\nonumber
\\
&&  \quad \quad \quad \quad \quad \quad \quad 
{\cal{R}} \, \,  = \,   \, d \, \tilde{\omega},
\end{eqnarray}
which is Cartan's curvature equation in 2 dimensions.


Up to now in
   Sect.~\ref{sec.curvature.from.Newton},
an LONB-rotation in 2 dimensions has been given by 
{\it one number},
the rotation angle. 
We now start using again the {\it matrix notation}
for infinitesimal LONB-rotations and Lorentz-boosts 
$\, \omega^{\hat{a}}_{\, \, \, \hat{b}} \,$
for a given displacement in the coordinate basis 
$\, \partial_{\nu} 
\equiv \bar{e}_{\nu}, \, $
which gives {\it Cartan's connection 1-form}
$ \, (\omega^{\hat{a}}_{\, \, \, \hat{b}})_{\nu}.   $
Taking the exterior derivative with $\, \partial_{\mu} \, $
as in 
   Eq.~(\ref{Cartan.2form.comp.2dim}), 
but now in the notation of an infinitesimal 
LONB-transformation-matrix, 
gives,
\begin{eqnarray}
\mbox{Cartan's curvature 2-form} 
\, \, &=& \,\,({\cal{R}}^{\hat{a}}_{\, \, \, \hat{b}})_{\mu \nu}.
\nonumber
\\
\mbox{LONB transformation indices}
\, \, &=& \, \, [\, \hat{a}, \, \hat{b} \, ],
\nonumber
\\
\mbox{plaquette for displacements in coord.} 
\, \, &=& \, \, [\, \mu, \, \nu \, ].
\end{eqnarray}

This completes our derivation of 
{\it Cartan's curvature equation} in (1+1)-spacetime and in 2-space,
which started with the total rotation angle of LONBs around a closed curve,
   Sec.~\ref{sect.Cartan.LONB.rot.closed.curve},
\begin{eqnarray}
({\cal{R}}^{\hat{a}}_{\, \, \, \hat{b}})_{\mu \nu}
\, &=&  \, (\, d \, \omega^{\, \hat{a}}_{\, \, \, \hat{b}})_{\mu \nu},
\nonumber
\\
{\cal{R}}^{\hat{a}}_{\, \, \, \hat{b}}
\, &=&  \, d \, \omega^{\, \hat{a}}_{\, \, \, \hat{b}}.
\end{eqnarray}
%

\newpage

\boldmath
\subsection{Exact Ricci curvature $R^{\hat{0} \hat{0}}$
from 
\protect\\ 
relative acceleration of quasistatic particles 
\protect\\
in (1+1)-spacetime}
\unboldmath

We prove the crucial new theorem stated in the title of this section, 
valid in {\it arbitrarily strong gravitational fields}.
As a preparation, we start  
with  (1+1)-dimensional spacetime in this subsection .

We compute the curvature in
(1+1)-spacetime with our spacetime slicing $\Sigma_t$ and our radially
parallel LONBs.
We also need  a spatial coordinate $\, P \Rightarrow x_P. \,$
We choose the Riemann normal 1-coordinate $x$ with $g_{xx} \equiv 1,$
i.e. the coordinate $x_P$ is equal to 
the measured distance from the observer to $P$ on $\Sigma_t$.

%
\begin{itemize}
\item
For {\it curvature} computations in this first paper,
we consider an {\it inertial primary observer}.
This is sufficient to derive Einstein's $R^{\hat{0} \hat{0}}$ equation
from Newtonian experiments.
\end{itemize}
In a second paper, 
spacetime curvature will be computed 
for a non-inertial observer. 

We choose the spacetime {\it coordinate-plaquette} $\, [\partial_{t},
\, \partial_{x} ] \,$ 
formed 
on one side by the worldline of the primary observer at $x = 0.$

(1) For a displacement along a worldline of an {\it inertial primary observer}
at $\, r = 0$, 
\begin{eqnarray}
&& \mbox{along inertial worldline}: 
\quad \quad [(\omega_{\hat{r} \hat{t}})_t]_{x = 0} 
\, \, = \, \, 0.
\nonumber
\end{eqnarray}

(2) With {\it LONBs spatially parallel},
the LONB Lorentz-boost angles {\it vanish} in the $x$ direction, $t = $ fixed,
\begin{eqnarray} 
\mbox{along spatial geodesics}: \, \, &&\, \, (\omega_{\hat{x} \hat{t}})_x 
\, \, = \, \, 0.
\nonumber
\end{eqnarray}

(3) For an {\it auxiliary observer} at  $ x = $ fixed and infinitesimal,
the LONB-Lorentz-boost angle for a displacement along his worldline,
i.e. the {\it relativistic} connection coefficients 
$\, (\omega_{\hat{a} \hat{b}})_{0},  \,$
are determined {\it exactly}
by experiments with {\it quasistatic inertial}  test particles
initially at rest relative to the auxiliary observer,
as in Eqs.~(\ref{def.E}, \ref{E.Ricci}),
\begin{eqnarray}
\mbox{along worldline} &&  \delta x  \, \, \mbox{constant:}
\nonumber
\\
 (\omega_{\hat{x} \hat{t}})_t  \,   
&=& \, - \, 
E_x^{(\rm g)} \, \,
\nonumber
\\ 
&=&  \, - \, 
(\frac{dv}{dt})_{\rm inertial\, particle}^{\rm quasistatic}
\, \, = \, \, - \, a_x.
\nonumber
\end{eqnarray}
With our choice of the spatial coordinate,
the {\it coordinate} distance $\, x \, $ 
is the {\it measured} geodesic distance.
Around the infinitesimal $[\, t, x\, ]$-plaquette of the observer,
the {\it coordinate} time $t$ is the time {\it measured} by the observer, 
to first order in $x$. 
The measured velocity of quasistatic particles is $\, v = dx/dt.$

The {\it relative (tidal) acceleration} 
of two neighbouring inertial quasistatic  test particles,
one constantly at rest relative to the primary inertial observer,
the other one {\it starting} at rest relative 
to the primary and to the auxiliary observers,
%
\begin{eqnarray}
\partial_x [(   \omega_{\hat{x} \hat{t}}  )_t] \,    &=& \,
- \, \mbox{div} \, \vec{E}_{\rm g} \, \, 
= \, - \, 
\partial_x  \, [(\frac{dv}{dt})_{\rm freefalling}^{\rm quasistatic}],
\label{relat.accel.without.curvature} 
\end{eqnarray}
where $\, \partial_x E_x^{(\rm g)} = \mbox{div} \, \vec{E}_{\rm g}, \,$
because the LONBs are radially parallel at {\it all} times.

The space-time curvature is obtained
by considering 
the {\it deficit Lorentz-boost angle} $\delta$ of LONBs  
around our infinitesimal 
$[\, t, x\, ]$-plaquette, 
\begin{eqnarray}
\delta
&=& - \oint_{[x,t]} \, (\omega_{\hat{x} \hat{t}})_{\mu} \, dx^{\mu},
\label{deficit.Lorentz.boost.angle.2}
\end{eqnarray}
where the displacements in $x$ do not give a contribution, 
since the LONBs are radially parallel, 
and the displacement along the 
primary observer's worldline at $x = 0$
does not give a contribution,
since the primary observer is inertial, hence 
the LONBs along his worldline are self-parallel.

The deficit Lorentz-boost  angle 
   $\delta $ 
per measured plaquette area
is per definition 
Cartan's (1+1) Riemann curvature  
$({\cal{R}}^{\hat{t}}_{\, \, \,  \hat{x}})_{tx}$.
With coordinate indices equal to LONB indices at $P_0$,
$\bar{e}_{a} = \bar{e}_{\hat{a}}$,
the contraction of the two spatial indices
gives the Ricci component $\, R^{\hat{t}}_{\, \, \,  \hat{t}}$ 
\begin{eqnarray}
&&\mbox{inertial primary observer, radially parallel LONBs}
\nonumber
\\
&&R^{\, \hat{t}}_{\, \, \,  \hat{t}} =      
({\cal{R}}^{\hat{t}}_{\, \, \,  \hat{x}})_{\hat{t} \hat{x}}
=  - \mbox{div} \, \vec{E}_{\rm g}  
=    
- \partial_x   (a_x)_{\rm freefalling}^{\rm quasistatic},
\label{R.zero.zero.tidal.acc.div.E.2}
\end{eqnarray}
%

The relationship between curvature and measurements by 
{\it non-inertial observers} will be given in a second paper.

%
\begin{itemize}
\item
One can measure the curvature $\, R^{\hat{0} \hat{0}} \,$ 
{\it exactly} 
in arbitrarily {\it strong} gravitational fields 
with test-particles
which are {\it quasi-static} relative to the {\it inertial} 
observer with $\, \bar{u}_{\rm obs} = \bar{e}_{\hat{0}} \, $
as shown in
   Eq.~(\ref{R.zero.zero.tidal.acc.div.E.2}).
\end{itemize}
%

\subsection{Cartan's curvature equation 
\protect\\
in more than two dimensions}

For Cartan's curvature equation in three and more dimensions, 
we need the {\it antisymmetric product} of two 1-forms 
$\tilde{\sigma}$
and 
$\tilde{\rho}$, 
called the {\it exterior product} and denoted by a {\it wedge}, 
$\tilde{\sigma} \wedge \tilde{\rho}$,
\begin{eqnarray}
\mbox{exterior product:} \quad 
[\, \tilde{\sigma} \wedge \tilde{\rho}   \, ]_{\mu \nu}
\, &\equiv& \, \sigma_{\mu} \, \rho_{\nu} - \sigma_{\nu} \, \rho_{\mu}.\quad
\end{eqnarray} 

The {\it Hodge-star-dual}  of the exterior product of two 1-forms
in three spatial dimensions,
$[ \, \star \, (\, \tilde{\sigma} \wedge \tilde{\rho} \, ) \, ]_{i},  $ 
is equal to the {\it vector product}
$\, [ \, \vec{\sigma} \times \vec{\rho} \, ]_{i}   \, $,
\begin{eqnarray}
[ \, \star \, (\, \tilde{\sigma} \wedge \tilde{\rho} \, ) \, ]_{\hat{i}} 
\, &=& \, 
\varepsilon_{\hat{i} \hat{j} \hat{k}} \, \, \sigma_{\hat{j}} \, \, \rho_{\hat{k}} 
\, \, = \, \, [\, \vec{\sigma} \times \vec{\rho}  \, ]_{i}.
\label{exterior.product.vector.product}
\end{eqnarray}
We omit all equations which are not needed for Cartan's curvature equation.

In dimensions $n$ higher than 2,   
a connection {\it Lorentz transformation} 
of {\it LONBs} is given by a (n x n)-matrix.
For a {\it coordinate-displacement} $\mu$,
it is given by {\it Cartan's connection-1-form}
$\, (\omega^{\hat{a}}_{\, \, \, \hat{b}})_{\mu}$.

The curvature, 
the deficit Lorentz transformation under a displacement
around the coordinate plaquette $\, [\, \mu, \, \nu \, ],   \,$
divided by the plaquette area, 
is given by, 
\begin{eqnarray}
&&  
({\cal{R}}^{\hat{a}}_{\, \, \, \hat{b}})_{\mu \nu}:
\label{notation.Riemann.tensor}
\\
\, \hat{a}, \, \hat{b}   \,  
&=& \mbox{LONB indices of  deficit Lorentz-transf.},
\nonumber
\\ 
\, \mu, \, \nu \,  &=& \mbox{coord. plaquette displacement-indices}.
\nonumber
\end{eqnarray}
Dropping the displacement indices around the plaquette
gives the {\it curvature-2-form} 
$\, {\cal{R}}^{\hat{a}}_{\, \, \, \hat{b}}.\,$

In more than two dimensions, 
Cartan's curvature formula has an {\it extra term},
which {\it vanishes} for our primary-inertial-observer adapted LONBs.~---
We report
Cartan's curvature equation from
   \cite{MTW},
\begin{eqnarray}
{\cal{R}}^{\hat{a}}_{\, \, \, \hat{b}} \, &=& \, 
d \, \omega^{\, \hat{a}}_{\, \, \, \hat{b}} \, \, + \, \, 
\omega^{\, \hat{a}}_{\, \, \, \hat{s}} \wedge 
\omega^{\, \hat{s}}_{\, \, \,  \hat{b}}.
\label{Cartan.curvature.eq.general}
\end{eqnarray}
The {\it second term vanishes}
on the worldline of a {\it inertial primary observer} 
with his {\it radially parallel LONBs},
because {\it all  connection coefficients vanish on his worldline}.

\boldmath
\subsection{Exact Ricci curvature $\, R^{\hat{0} \hat{0}} \, $ from
\protect\\
relative acceleration 
of quasistatic particles
\protect\\
in (3+1) space-time} 
\unboldmath

%

{\it Along} the worldline of a primary observer, 
his LONBs and our coordinate bases are identical, 
$\bar{e}_{\hat{a}} = \bar{e}_{a}.$

{\it Away} from the worldline of a primary inertial observer,
$\bar{e}_{\hat{a}} = \bar{e}_{a}$ continues to hold 
to {\it first order} in $r$
for our choice of slicing $\Sigma_t,$
 LONBs, and 3-coordinates.

In curved (3+1) spacetime, 
a primary inertial observer 
measures the
{\it spherical average} $\, <...>_{\rm spherical} \,$ of the
{\it radial accelerations} $\, a_{\rm radial} = (dv_r/dt) \,$  
of {\it freefalling} test-particles initially at measured radial distance $r$.
The test-particles at $r$ can be chosen 
{\it quasi-static} relative to the primary observer at $r = 0$,
and they will remain quasistatic within an infinitesimal time-interval. 
The {\it spherical average} is equal 
to the {\it average} over the {\it three principal directions} 
$(\vec{e}_{\hat{x}}, \, \vec{e}_{\hat{y}}, \, \vec{e}_{\hat{z}})$.
Hence,
the generalization of 
     Eq.~(\ref{R.zero.zero.tidal.acc.div.E.2})
to (3+1) spacetime is,
\begin{eqnarray}
&& \mbox{for inertial primary observer:}
\nonumber
\\ 
( R^{\, \hat{0} \hat{0}} )_{\rm exact} &=&  
 (- \partial_r 
<a_{\rm radial}^{\rm relative}>_{\rm spherical})^{\rm freefalling}_{\rm
  quasistatic}
\nonumber
\\ 
&=& (\mbox{div} \, \vec{E}_{\rm g})_{\rm inertial \, primary \, observer}.
\label{3d.R.zero.zero.tidal.acc.div.E.2}
\end{eqnarray}
%
Both equations are wrong for: 
\begin{itemize}
\item 
a non-inertial primary observer,
\item  
secondary observers (to obtain the relative acceleration)
whose LONBs are not radially parallel with the LONBs of the primary observer.
\end{itemize}
Conclusion:
\begin{itemize}
\item
For a {\it freefalling-nonrotating primary observer} with 
$\, \bar{u}_{\rm obs} = \bar{e}_{\hat{0}},$
the {\it exact} Ricci component
    $ \, R_{\hat{0} \hat{0}} \, $ 
of general relativity for arbitrarily {\it strong gravity} fields is
\begin{center} 
{\it unambigously determined}
\end{center}
by measuring the relative acceleration
of neighbouring freefalling test-particles which are
{\it quasistatic} relative to the observer.

$R_{\hat{0} \hat{0}} \, $ is {\it linear} in the gravitational fields.
\end{itemize}
Observers which are not freefalling-nonrotating will be treated in a separate
paper.~---
For a uniformly rotating-freefalling observer, 
there is only {\it one} fictitious force, the centrifugal force,
causing the {\it radial acceleration} of quasi-static 
test-particles,
\begin{eqnarray} 
\vec{a}_{\rm centrifugal} 
&=& [ \, \vec{\Omega}_{\rm obs} \times 
  [ \, \vec{r} \times \vec{\Omega}_{\rm obs} \, ] \, ]
\nonumber
\\
&=& \frac{1}{4} \, [ \, \vec{B}_{\rm g} \times 
    [ \, \vec{r} \times \vec{B}_{\rm g} \, ] \, ].
\end{eqnarray}
The radial derivative of the centrifugal acceleration at $r=0$
gives an additional term for $\, R^{\hat{0} \hat{0}}, \, $ 
\begin{eqnarray}
 R^{\hat{0} \hat{0}}_{\rm extra} &=& 
 (- \partial_r 
<a_{\rm radial}^{\rm relative}>^{\rm extra}_{\rm spherical})^{\rm freefalling}_{\rm
  quasistatic}
\nonumber
\\
&=& 
- \, \mbox{const.}  \, \vec{B}^2_{\rm g}.
\label{centrifugal.term.in.curvature}
\end{eqnarray}

\boldmath
\subsection{Einstein's $R^{\hat{0} \hat{0}}$ equation
\protect\\
for nonrelativistic matter-sources
\protect\\
identical with
\protect\\ 
Newton's equation of relative acceleration} 
\unboldmath

The exact equality  between 
the space-time {\it curvature} $R^{\,\hat{0} \hat{0}} (P)$
and the {\it relative acceleration} of freefalling particles,
spherically averaged,
for a {\it freefalling-nonrotating observer} with worldline through $P$
and $\, \bar{u}_{\rm obs} (P) =\bar{e}_{\hat{0}} (P), \, $ 
    Eq.~(\ref{3d.R.zero.zero.tidal.acc.div.E.2}),
gives the proof that,
%
\begin{itemize}
\item
In arbitrarily strong gravitational fields,
{\it Newton's} law for 
acceleration of 
freefalling particles at $\delta r$,  
measured relative to a freefalling-nonrotating observer
and spherically averaged,
%
%
\begin{eqnarray}
<a_{\rm radial}>^{\rm angular \, av.}_{\rm inertial \, obs.} \,  
&=&  \, - \, G_{\rm N} \, \, M_{\delta r} 
\, \, (\delta r)^{-2},
\end{eqnarray}
%
%
\begin{center}
is {\it explicitely identical} 
\end{center}
%
with {\it Einstein's} $R^{\, \hat{0} \hat{0}}$ equation for
{\it source-matter non-relativistic} relative to this inertial observer,
\begin{eqnarray}
R^{\hat{0} \hat{0}} &=& 4\pi G_{\rm N} \, \rho_{\rm mass}.   
\end{eqnarray}
\end{itemize}
%

Only {\it after choosing} 
an {\it inertial primary observer} 
and  
{\it LONBs radially parallel}, 
is $\, \mbox{div} \, \vec{E}_{\rm g} \,$ 
uniquely defined for a given physical situation.
With these two choices
and only with these choices,
we obtain,
\begin{eqnarray}
R^{\hat{0} \hat{0}} 
\, = \, - \, 
(\,\mbox{div} \, 
\vec{E}_{\rm g} \, )^{\rm LONBs \,  rad. parallel}_{\rm inertial \, primary \, obs.}
\, = \, 4 \, \pi \, G_{\rm N} \, \rho_{\rm mass}
\label{R.0.0.eq.source.nonrelat}
\end{eqnarray}
%

The explicit field equation
         for {\it non-inertial observers} 
is {\it bilinear} in the gravitational fields 
$(\vec{E}_{\rm g}, \, \vec{B}_{\rm g})$. 
These explicit field equations
will be given  in a second paper.

For $\, R^{\hat{0} \hat{0}}, \,$
a tensor component, 
it is {\it irrelevant}, 
whether the  primary observer is {\it inertial} or {\it non-inertial},
and it is irrelevant, whether we choose LONBs 
away from the primary worldline radially parallel or not
(which dictates the choice for auxiliary observers).~---
But $\, \mbox{div} \, \vec{E}_{\rm g}  \,$ involves Ricci connections,
{\it all depends} on inertial versus non-inertial primary observer
(field equation linear in $\vec{E}_{\rm g}$ versus bilinear) 
and it depends on LONBs radially parallel or not
(two fields, $\, \vec{E}_{\rm g} \, $ and $ \, \vec{B}_{\rm g}, \,  $              
versus more than a hundred connection coefficients).

It has often been  emphasized that
a fundamental difference between
general relativity and Newtonian physics
is the {\it non-linearity} of Einstein's equations
versus
the {\it linearity}
of the Newton-Gauss field equation
$\, \mbox{div} \, \vec{E}_{\rm g} = - 4 \pi G_N \rho_{\rm mass}.$~---
Nothing could be farther from the truth: 
\begin{itemize}
\item  {\it Einstein's}  
       $R^{\hat{0} \hat{0}} (P)$ equation
       with nonrelativistic source-matter 
       and the gravitational field equation  
       of  19th-century 
       {\it Newton-Gauss} physics 
       are 
\begin{center}
   {\it explicitely identical.}
\end{center} 
       Both field equations are:
    \begin{enumerate}
       \item
         {\it linear} 
         for a {\it freefalling-nonrotating observer} 
         with worldline through $P$,
         with $\bar{u}_{\rm obse} = \bar{e}_{\hat{0}}$,
         and with our radially parallel LONBs,
       \item
         {\it nonlinear}
         for {\it non-freefalling and/or rotating observers} 
         with worldlines through $P$,
         because fictitious forces  are
         equivalent to gravitational forces,
             Sec.~\ref{sec.equiv.th}, 
         and e.g. the centrifugal term is quadratic
         in $\vec{B}_{\rm g},$
            Eq.~(\ref{centrifugal.term.in.curvature}).
     \end{enumerate}
%

%
%
%
%
%
%
\end{itemize}
%

\subsection{From nonrelativistic source-matter 
to the full Einstein equations} 

Using,
\begin{enumerate}
\item Lorentz covariance,
\item the contracted second Bianchi identity to satisfy the covariant
  conservation of the energy-momentum-stress tensor,
\end{enumerate}
one directly obtains Einstein's equations, as explained in all textbooks,

\begin{eqnarray}
G^{\hat{a} \hat{b}} \, &=& \, 8 \, \pi \, G_{\rm N} \, T^{\hat{a} \hat{b}},
\label{final.Einstein.eq}
\\
G^{\hat{a} \hat{b}} \, &\equiv& \, 
R^{\hat{a} \hat{b}} \, - \frac{1}{2} \, \eta^{\hat{a} \hat{b}} \, R.
\end{eqnarray}
%
This completes our derivation of Einstein's equations of general relativity
from Newtonian experiments on relative acceleration.

\subsection{Comparison with ``heuristic derivations''
\protect\\ 
of Einstein's equations}

There are two important differences between our paper and the literature:

The first crucial difference is 
that our paper gives a rigorous derivation of
Einstein's $\, R^{\hat{0} \hat{0}} \, $ equation from nothing more than
Newtonian phsics and Einstein's concept of inertial motion,
while   
the literature 
only gives ``heuristic derivations'' of Einstein's equations.

Straumann
  \cite{Straumann.heuristic},
for his Sec.~3.2.1 
gives  the title:  Heuristic ``Derivation'' of the Field Equations
(`` ... `` marks by Straumann).  

Wald
  \cite{Wald.heuristic},  
in contrast to a rigorous derivation, 
writes in Section 4.3, 
``a clue is provided'',
``the correspondence suggests the field equation'',
but Wald gives no equation for the correspondence.

Weinberg
\cite{Weinberg.guess} in Sec.~7.1
takes the
"weak static limit", makes a "guess", and argues with "number of derivatives".

Misner, Thorne, and Wheeler
   \cite{MTW.six.routes} 
give
"Six Routes to Einstein's field equations" in Box 17.2:
(1)~to model geometrodynamics after electrodynamics, 
(2)~to take the variational principle with 
only a scalar linear in second derivatives
of the metric and no higher derivatives, 
(3)~``again electromagnetism as a model'',
(4)~superspace,
(5)~field equation for spin 2,
(6)~Sakharov's view of gravitation as an elasticity of space.

In contrast, this paper gives 
a rigorous derivation of Einstein's equations from Newtonian physics,
not heuristic arguments.

The second crucial difference between the literature and our paper:
For the $\, R^{\hat{0} \hat{0}}  \,$ field equation at a given point $P$,
one can distinguish the matter-source ``non-relativistic versus
relativistic''.
But the term ``Newtonian limit'' 
is meaningless for the  $\, R^{\hat{0} \hat{0}}  \,$ field equation,
because this Einstein equation is {\it identical} to 
the 19th-century Newtonian field equation: 
both equations are identical and {\it linear} 
for an {\it inertial observer} with worldline through $P$,
both equations  are {\it identical} and {\it non-linear} 
for {\it non-inertial observers}. 

\subsection{Outlook}

In this paper, 
the field equation for 
$\, R^{\, \hat{0} \hat{0}} \, (P) \, $
has been treated 
only  
for an {\it inertial primary observer} 
with worldline through $P$ and with
$\, \bar{u}_{\rm obs} \, (P) = \bar{e}_{\hat{0}} \ (P)$.
This  field equation of Einstein 
is {\it linear} in the gravitational field $\vec{E}_{\rm g}$,
and it is explicitely identical with the Newton-Gauss field equation,
$\, \mbox{div} \, \vec{E}_{\rm g} = - 4 \pi G_{\rm N} \, \rho_{\rm mass}.$

A first extension:
The field equation for  {\it non-inertial observers} will be given
in a companion paper.
Again, the 19th-century Newtonian equation 
and Einstein's $\, R^{\hat{0} \hat{0}} $ equation
for non-relativistic source-matter
are {\it exactly} and {\it explicitely identical} for arbitrarily strong
gravitational fields.
But for {\it non-inertial observers},
both the field equation of 19th-century Newtonian physics
 and Einstein's  $\, R^{\hat{0} \hat{0}} $
equation are {\it bilinear}
in the gravitational fields.
%
%

A second extension:
      One asks the question, 
      can this linearity for inertial observers be extended 
      to an equation and its solution
      all over the universe?~---
      In the paper  
            \cite{Texas.2013},
      we have  derived and 
      solved {\it exactly}
      the {\it linear} field equation for 
      Einstein's angular-momentum constraint at any point $P_0$
      from the angular-momentum input 
      of an arbitrary distribution of {\it matter}
      on the past light-cone of $P_0$.
      The output is $\, \vec{B}_{\rm g} \,$ at $P_0$,
      given as a simple explicit {\it linear retarded integral}
      over the matter-angular-momentum input
      on the past light-cone of $P_0$ 
      back to the big bang.
\bibliography{paperY.bib}

\end{document}